\documentclass [11pt]{article}

\usepackage{amsmath,amssymb,epsfig,cite,color,verbatim}
\usepackage{graphics}

\numberwithin{equation}{section}

\title{{\bf Quantum Phases of Yang-Mills Matrix Model Coupled to Fundamental Fermions}}
\author{Mahul Pandey\footnote{mahul@cts.iisc.ernet.in} \,and Sachindeo Vaidya\footnote{vaidya@cts.iisc.ernet.in} \\
\begin{small}{\it Centre for High Energy Physics, Indian Institute of Science, Bangalore, 560012, India}
\end{small}}

\date{}

\begin{document}
\maketitle
\begin{abstract}
By investigating the $SU(2)$ Yang-Mills matrix model coupled to fundamental fermions in the adiabatic limit, 
we demonstrate quantum critical behaviour at special corners of the gauge field configuration space. The 
quantum scalar potential for the gauge field induced by the fermions diverges at the corners, and is 
intimately related to points of enhanced degeneracy of the fermionic Hamiltonian. This in turn leads to 
superselection sectors in the Hilbert space of the gauge field, the ground states in different sectors being 
orthogonal to each other.  As a consequence of our analysis, we show that 2-color QCD coupled to two 
Weyl fermions has three quantum phases. When coupled to a massless Dirac fermion, the number of 
quantum phases is four. One of these phases is the color-spin locked phase.
\end{abstract}

\section{Introduction}

The nature of the ground state of quantum chromodynamics (QCD) is still in the process of being understood, and 
is a subject of intense theoretical and numerical activity. QCD, or more generally,  non-Abelian Yang-Mills theory 
coupled to fundamental fermions (or quarks) displays a diverse variety of phases, even at zero temperature. That many of these phases occur at 
zero temperature strongly suggests that transitions between these phases are driven by quantum rather than statistical fluctuations. Many of 
these phases are spatially homogeneous, characterised by fermion condensates of uniform density, as well as uniform chromo-electric and/or 
chromo-magnetic fields (see for instance \cite{Alford:2007xm}).

This suggests that approximating the full theory by degrees of freedom that are spatially homogeneous can provide better insight into this 
phase structure. A more precise formulation of such an approximation is in terms of a gauge matrix model, that say, corresponds to reducing the full 
Yang-Mills theory on $S^3 \times \mathbb{R}$. Such matrix models are also interesting approximations of the full quantum field theory in their own right, capable of capturing many topological features and low-energy dynamics of the gauge fields. 
One such matrix model for $SU(N)$ Yang-Mills theory has been developed in \cite{Balachandran:2014iya,Balachandran:2014voa}, where the focus of investigation has been to study the nature of impure states in Yang-Mills theories. The gauge field is a rectangular matrix $M_{ia}$ with $i=1,2,3$ and $a=1,\cdots N^2-1$. In particular, the authors explicitly demonstrate the mixed nature of coloured QCD states as a consequence of the twisted nature of the QCD gauge bundle \cite{Singer:1978dk,Narasimhan:1979kf}. 

In this paper, we couple the $SU(2)$ matrix model of \cite{Balachandran:2014iya,Balachandran:2014voa} to fermions in 
the fundamental representation, and after a suitable rescaling of the gauge and fermionic variables, quantize the theory in the background-field approximation. 
As we shall see, such an approximation is appropriate 
for small values of Yang-Mills coupling $g$, which makes the contribution of the kinetic term of the Yang-Mills Hamiltonian much smaller 
than the potential as well as the fermionic terms. 
This situation is reminiscent of the Born-Oppenheimer (B-O) approximation 
in quantum molecular dynamics, where the atomic nuclei are slow degrees of freedom (precisely because the contribution of their 
kinetic energy to the total Hamiltonian is small), and the surrounding cloud of electrons the fast degrees of freedom. The dynamics of the nuclei is treated adiabatically, and a careful treatment of the B-O approximation leads to an adiabatic scalar potential induced in the space of slow variables \cite{berry,Bohm}, in addition to the well-known adiabatic Berry connection. 

Our strategy is to quantize the fermions in the background of the gauge fields in the adiabatic approximation, and solve for the exact
fermionic spectrum in terms of invariants of the matrix $M_{ia}$. The fermions in turn induce an effective scalar potential for 
the gauge fields,  whose singularities we examine in some detail. We show that the scalar potential diverges at certain edges and corners of the gauge 
field configuration space. This divergence is rather easy to understand: it corresponds to places where the degeneracy of the 
fermion spectrum changes, or loosely speaking, at points of fermionic level crossing. The induced scalar potential can also be computed directly 
at the degenerate point(s) of the fermionic spectrum, and is perfectly well-defined. This leads to a remarkable situation for the Hilbert space of the 
gauge variables: it gets divided into different superselection sectors, which may be interpreted as 
different quantum phases.

The singularity structure of the effective potential allows us to identify several quantum phases in $SU(2)$ Yang-Mills theory coupled 
to fermions with one flavor. An immediate corollary of our work is our identification of a {\it color-spin} locked phase analogous to the one predicted 
in 3-color QCD \cite{Schafer:2000tw}.

The article is organized as follows. In Section 2, we introduce our model, describe the Hamiltonian and its physical and 
gauge symmetries. In Section 3 we quantize the model in the B-O approximation, and show how the effective scalar potential emerges naturally as 
a consequence of the adiabatic approximation. We also discuss the issue of implementing Gauss' law in the B-O scheme, and its implications 
for possible breaking of gauge invariance. In Section 4, we compute the fermion spectrum and study its degeneracy structure. The fermion 
spectrum, through its dependence of the gauge field, allows us to identify certain classical gauge configurations as {\it edges} and 
{\it corners} of the gauge configuration space. For the case of 
$SU(2)$, we discover, in passing, some unexpected inequalities obeyed by all $3 \times 3$ real matrices. We also show that at these edges/corners, there is an enhancement of symmetry of the fermion Hamiltonian. In Section 5 we discuss the properties of the adiabatic connection at the edges and corners of the gauge configuration space. In Section 6, we investigate the scalar potential induced 
by the 
two-fermion state, and show that it diverges as one approaches points of enhanced fermion degeneracy.  We also compute the scalar potential 
directly at the edges and corners, and see that it is perfectly well-behaved. The emergence of superselection sectors for gauge field dynamics is discussed in Section 7. Section 8 presents an analogous discussion for massless Dirac fermions, with similar conclusions. 
Our conclusions are presented in Section 9.

Finally, a word about the use of the phrase ``quantum phases". Many of the models studied to date that display quantum phase transitions have tunable couplings in the Hamiltonian. The adiabatic scalar potential has singularities which carry information about putative locations of quantum phase transitions \cite{Sachdev,Zanardi,Dutta} in the space of couplings. These are also characterized by a non-analytic behaviour of the spectrum as one approaches the critical point(s) (or critical regions) in the coupling constant space, and the ground states on either sides of the critical point are orthogonal to each other. All these features, namely, divergence of the scalar potential, non-analyticity of the spectrum, and orthogonality of the ground state across critical points (or regions) are present in our situation. The main difference, as we see it, is the absence of tunable couplings in our model: more precisely, the couplings, rather than being externally tunable, have their own quantum dynamics. This leads to the
emergence of superselection sectors in the 
Hilbert space for gauge variables. We interpret these superselection sectors as different quantum phases of the theory.

\section{The Matrix Model for Weyl Fermions}

The $SU(2)$ matrix model of \cite{Balachandran:2014iya,Balachandran:2014voa} is obtained by starting with the Yang-Mills theory on $S^3 \times \mathbb{R}$ and isomorphically mapping the spatial $S^3$ to $SU(2)$, with the three left-invariant vector fields $X_i$ on $S^3$ identified with 
$\textstyle{\frac{\tau_a}{2}}$ of $SU(2)$. Consider an arbitrary left invariant form $\Omega$ on $SU(2)$:

\begin{equation}
\Omega = {\rm Tr} \left( \frac{\tau_a}{2} u^{-1} d u\right)M_{ab} \frac{\tau_b}{2}, \quad u \in SU(2).
\end{equation}

The Hermitian gauge field is simply the pullback of $\Omega$ under the isomorphic mapping of the spatial $S^3$ to $SU(2)$:
\begin{equation}
A_i = i\Omega(X_i) =M_{ia} \frac{\tau_a}{2},
\end{equation}
and 
\begin{equation}
A_0 = M_{0a} \frac{\tau_a}{2}.
\end{equation}
The curvature $F_{ij}$ corresponding to this $A_i$ is obtained by the pull-back of the Maurer-Cartan form $d \Omega + \Omega \wedge \Omega$ to 
the spatial $S^3$:
\begin{eqnarray}
F_{ij} &=& (d \Omega + \Omega \wedge \Omega)(X_i, X_j), \\
F_{ij}^a &=& -\epsilon_{ijk}M_{ka}+f_{abc}M_{ib}M_{jc}.
\end{eqnarray}

The chromoelectric field $E_i^a = F_{0i}^a = \dot{M}_{ia}+f_{abc}M_{0b}M_{ic}$ and the chromomagnetic field $B_i^a = \frac{1}{2}\epsilon_{ijk}F_{jk}^a$ give us the Lagrangian for the matrix model
\begin{equation}
L_{YM}= -\frac{1}{4g^2}F_{\mu\nu}^aF^{a\mu\nu}=\frac{1}{2g^2} \left( E_i^a E_i^a -B_i^a B_i^a \right)
\end{equation}
Fermions can be introduced by minimal coupling \cite{Sen:1985dc}:
\begin{equation}
L=-\frac{1}{4g^2}F_{\mu\nu}^aF^{a\mu\nu}
+\left( i\lambda^\dagger_{A} \bar{\sigma}^\mu (\mathcal{D}_\mu\lambda)_{A}
+\lambda^\dagger _{\alpha A}\lambda_{\alpha A}\right), \quad {\rm where} \quad \sigma^\mu=(\mathbf{1},\sigma^i),\quad \bar{\sigma}^\mu=(\mathbf{1},-\sigma^i)
\end{equation}
and
\begin{eqnarray}
(\mathcal{D}_0 \lambda)_A=\partial_0 \lambda_A+\frac{i}{2}M_{0c}(\tau_c)_{AB}\lambda_B, \quad\quad
(\mathcal{D}_i \lambda)^a=\frac{i}{2}M_{ic}(\tau_c)_{AB}\lambda_B.
\end{eqnarray}
The term $\lambda^\dagger \lambda$ in ($2.7$) comes from the curvature of $S^3$.

For our discussion, it is useful to rescale the fermionic variables as $\lambda \rightarrow g \lambda$. The Lagrangian then becomes 
\begin{equation}
L=-\frac{1}{4g^2}F_{\mu\nu}^aF^{a\mu\nu}
+\frac{1}{g^2}\left( i\lambda^\dagger_{A} \bar{\sigma}^\mu (\mathcal{D}_\mu\lambda)_{A}
+\lambda^\dagger _{\alpha A}\lambda_{\alpha A}\right), 
\label{YMFLagrangian}
\end{equation}

For the $SU(2)$ model, the gauge variables are $3\times 3$ real matrices depending only on time:
\begin{equation}
M_i(t)=M_{ia}(t)\frac{\tau^a}{2} \quad\quad a=1,2,3,
\end{equation}
and $\tau_a$ are the usual Pauli matrices. The fermion field $\lambda\equiv \lambda_{\alpha A}$ also depend 
only on time.

Under gauge transformations, the gauge variables transform in the adjoint, and the fermions in the 
fundamental representation of the gauge group:

\begin{equation}
M_{ia}\rightarrow S(g)_{ab}M_{ib}, \quad \lambda_{\alpha A}\rightarrow s_{AB}(g)\lambda_{\alpha B};\quad A,B=1,2 
\quad {\rm and} \quad g\in SU(2).
\end{equation}

Under rotations,
\begin{equation}
M_{ia}\rightarrow R_{ij}M_{jb}, \quad \lambda_{\alpha A}\rightarrow r_{\alpha\beta}(R)\lambda_{\alpha B};\quad 
\alpha, \beta =1,2 \quad {\rm and} \quad R\in SO(3).
\end{equation}

The conjugate momenta are
\begin{equation}
\Pi_{ia}=\frac{\partial L}{\partial \dot{M}_{ia}}=\frac{1}{g^2}F_{0i}^a, \quad\quad
\Pi_{\alpha A}=\frac{\partial L}{\partial \dot{\lambda}_{\alpha A}}=
\frac{i}{g^2}\lambda^\dagger_{\alpha A}.
\end{equation}

Then, the Hamiltonian works out to be
\begin{eqnarray}
H' &=& \Pi_{ia}\dot{M}_{ia}+\Pi_{\alpha A}\dot{\lambda}_{\alpha A}-L, \\
&=& H+M_{0a}G^a, \quad {\rm where}\\
H &=& \frac{g^2}{2}\Pi_{ia}\Pi_{ia}+\frac{1}{4g^2}F_{ij}^a F_{ij}^a-\frac{1}{g^2}\lambda^\dagger_A\lambda_A-\frac{1}{2g^2}(\tau_b)_{AC}\bar{\lambda}^A\bar{\sigma}^i\lambda^C M_{ib}, \quad {\rm and} \label{hamonshell1} \\
G^a &=& \epsilon_{abc}\Pi_{ib}M_{ic}-\frac{1}{2g^2}(\tau_a)_{AB}\lambda^{\dagger}_B\lambda_C. \label{gausslaw}
\end{eqnarray}
This is a constrained system since the momentum conjugate to $M_{0a}$ is zero. $M_{0a}$ acts as a 
Lagrange multiplier in $H'$, with its coefficient being the Gauss' law constraint.

%

To quantize the system, we impose the canonical commutation (and anti-commutation) relations
\begin{equation}
\left[M_{ia},\Pi_{jb}\right]=i\delta_{ij}\delta_{ab}, \quad\quad\{ \lambda_{\alpha A},\lambda^\dagger_{\beta B} \}=g^2\delta_{\alpha\beta}\delta_{AB}
\end{equation}
and demand that all physical states $|\Psi\rangle_{phys}$ be annihilated by the Gauss law:
\begin{equation}
G^a |\Psi\rangle_{phys} =0.
\end{equation}

Wavefunctions are sections of appropriate vector bundles built on the gauge configuration space ${\cal C}={\cal M}/SO(3)$, where ${\cal M}$ is the space of all $3\times3$ real matrices. ${\cal C}$ is generically a 6-dimensional manifold, 3 of which correspond to physical rotations. 

Quantization of this space is subtle, because the action of $SO(3)$ on ${\cal M}$ is not free: ${\cal C}$ is a stratified space. As we shall see, coupling fermions to the gauge field provides us with a refined tool to deal with the strata. In particular we shall see that a change in the fermion degeneracy is accompanied by a change in the stratum.

We end this section by making a brief comparison to the usual perturbative description of Yang-Mills coupled to fermions. It suffices to make this argument in flat space $\mathbb{R}^3 \times \mathbb{R}$, it carries over easily to $S^3 \times \mathbb{R}$. The Lagrangian has the familiar form
\begin{equation}
L=-\frac{1}{4}{\rm Tr}\,F_{\mu\nu} F^{\mu \nu} + {\rm Tr}\,\bar{\psi} \gamma^\mu(\partial_\mu -i g A_\mu)\psi, \quad F_{\mu \nu}^a = \partial_\mu A_\mu^a-\partial_\nu A_\mu^a -i g[A_\mu, A_\nu]^a.
\end{equation}
Perturbation theory in the coupling constant $g$ is then performed, the dynamical variables being $A_\mu$ and $\psi$. To obtain our Lagrangian (\ref{YMFLagrangian}), we rescale the dynamical variables as $M_{ia} = g A_i^a$ and $\lambda = g \psi$. This also explains intuitively why the Born-Oppenheimer quantization of the next section, although performed at small $g$, is different from the usual perturbative quantization. For small $g$, when $M_{ia}$ and $\lambda$ are $O(1)$, the corresponding perturbative variables take large values of $O(1/g)$. Thus the B-O treatment 
focusses on that sector of the theory which has large values for the chromo-electric and chromo-magnetic fields. This sector is usually difficult to access in standard perturbation theory.

\section{Quantization in the Born-Oppenheimer approximation}

The Hamiltonian (\ref{hamonshell1}) can be written as

\begin{equation}
H=H_{YM}+H_f
\end{equation}
where 
\begin{eqnarray}
H_{YM} &=& \frac{g^2}{2}\Pi_{ia}\Pi_{ia}+\frac{1}{4g^2}F_{ij}^a F_{ij}^a, \\
H_f &\equiv& \frac{1}{g^2} \left(-\lambda^\dagger_{\alpha A}\lambda_{\alpha A}-\frac{1}{2}(\tau_b)_{AC}\lambda^\dagger_{\alpha A} (\bar{\sigma}^i)_{\alpha \gamma}\lambda_{\gamma C} M_{ib} \right)
\end{eqnarray}
For small $g$, the gauge kinetic term $\frac{g^2}{2}\Pi_{ia}\Pi_{ia}$ is small compared to the other terms on $H$. 
In this regime of $g$, it is appropriate to quantize the theory in the B-O approximation: we first solve for the spectrum of the fermionic Hamiltonian, treating the Yang-Mills field as a background field, and then quantize the 
gauge field dynamics. 
A modern treatment of the B-O scheme has been discussed in \cite{Bohm}, and we will adapt it to our matrix model below.

The space of physical states is the direct product of the Hilbert spaces for the fast and the slow motion:
\[ \mathcal{H}=\mathcal{H}_{slow}\otimes \mathcal{H}_{fast}.\]
The Hamiltonian is
\begin{eqnarray}
H &=& \frac{g^2}{2}\Pi_{ia}\Pi_{ia}+\frac{1}{g^2}(V(M))+h(M), \\
h(M) &=& -\lambda^{\dagger}_{\alpha A}(H_f(M))_{\alpha A,\beta B}\lambda_{\beta B}
\label{hM}
\end{eqnarray}
where
\begin{equation}
V(M)=\frac{1}{4}(F_{ij}^a F_{ij}^a)
\label{V}
\end{equation}
and
\begin{equation}
(H_f(M))_{\alpha A,\beta B}=\frac{1}{g^2}\left(-\mathbf{1}-\frac{1}{2}\sigma_i\otimes \tau_a M_{ia}\right)_{\alpha A,\beta B}.
\label{Hf}
\end{equation}

The eigenvalue problem
\begin{equation}
H|\psi^E\rangle=E|\psi^E\rangle, \,\, |\psi^E\rangle \in \mathcal{H}_{slow}\otimes \mathcal{H}_{fast}
\end{equation}
is solved in the B-O approximation, by first solving for the spectrum of $h(M)$.

To proceed, one needs to define a complete set of basis vectors in the full Hilbert space. One choice is the standard "position" eigenvectors, which is a direct product of the Yang-Mills configuration space and the Fock space of the fermions.
\begin{eqnarray}
|M;\lambda_{\alpha_1 A_1},...,\lambda_{\alpha_r A_r}\rangle &=& |M\rangle\otimes|\lambda_{\alpha_1 A_1},...,\lambda_{\alpha_r A_r}\rangle, \label{basis} \\
|\lambda_{\alpha_1 A_1},...,\lambda_{\alpha_r A_r}\rangle &=& \lambda_{\alpha_1 A_1}^\dagger...\lambda_{\alpha_r A_r}^\dagger|0\rangle.
\end{eqnarray}

Here, $r$ denotes the fermion number. Since there is no observable that can change the fermion number, the complete basis (\ref{basis}) 
can be further decomposed into sets of states corresponding to one fermion ($|M;\lambda_{\alpha A}\rangle$), two fermions ($|M;\lambda_{\alpha A}\lambda_{\beta B}\rangle$) and so on.

However, the basis vectors better suited for the B-O treatment are the generalized eigenvectors $|M;n\rangle$ of $M$ and $h(M)$:
\begin{equation}
h(M)|M;n\rangle=\epsilon_n(M)|M;n\rangle, \quad n=1, \ldots, N.
\end{equation}
%
%
%
This new basis allows us to represent the basis (\ref{basis}) as a ''twisted" direct product
\begin{eqnarray}
|M;n\rangle &=& |M\rangle \tilde{\otimes} |n(M)\rangle, \quad {\rm where} \\
h(M)|n(M)\rangle &=& \epsilon_n(M)|n(M)\rangle \quad n=1, \ldots, N.
\label{heqn}
\end{eqnarray}
This is not an ordinary tensor product, since $|n(M) \rangle$ depends on $M$. 

If we now expand the energy eigenfunctions in this basis, we obtain
\begin{equation}
|\psi^E\rangle = \sum_n \int dM'|M';n\rangle \psi^E_n(M'), \quad \psi^E_n(M') \equiv \langle M';n|\psi^E\rangle.
\end{equation}
The full wave-function is
\begin{equation}
|\psi_E\rangle=\sum_r \int dM |M;\lambda_{\alpha_1A_1}...\lambda_{\alpha_rA_r}\rangle \psi^E_{\alpha_1A_1,...\alpha_rA_r}, \quad
\psi^E_{{\alpha_1 A_1,...\alpha_r A_r}}(M)=\langle M;\lambda_{\alpha_1 A_1...\alpha_r A_r}|\psi^E\rangle.
\end{equation}

Then it is easy to see that
\begin{equation}
\psi^E_{{\alpha_1 A_1...\alpha_r A_r}}(M)=\sum_nC^n_{{\alpha_1 A_1...\alpha_r A_r}}(M) \psi_n^E(M),\quad
%
C^n_{\alpha_1 A_1,...,\alpha_r A_r}\equiv \langle \lambda_{\alpha_1 A_1}...\lambda_{\alpha_r A_r} |n(M)\rangle
\label{Cn}
\end{equation}
Eq. (\ref{Cn}) suggests that $C^n$ are the energy eigenfunctions expressed in the "position" basis. As it turns out, they are the eigenvectors of 
the fermion Hamiltonian $H_f$ with eigenvalue $\epsilon_n$. We will make this more precise in the next section where we compute the 
fermion spectrum.

Once the spectrum $\epsilon_n$ has been determined, we find that the Schr\"{o}dinger's equation for the "slow" motion takes the form
\begin{equation}
\sum_m \left[\frac{g^2}{2}\sum_l (-i\delta^{nl}\partial_{ia}-A_{ia}^{nl})(-i\delta^{lm}\partial_{ia}-A_{ia}^{lm})+\delta^{nm}\left(\frac{1}{g^2}V(M) + 
\epsilon_n(M)\right)\right]\psi_m^E(M)=E\psi_n^E(M), 
\end{equation}

where
\begin{equation}
A_{ia}^{mn}\equiv i\langle n(M)|\partial_{ia}|m(M)\rangle.
\end{equation}
This is in general a set of $N$ equations and is exactly equivalent to the original $N$-body problem. We are however, interested in
the effect of the fermion(s) occupying only their ground state; i.e. $n$ restricted to the ground state of $h(M)$, labelled as $n=0$. The sum over 
$m$ is also restricted to $m=0$. Labelling the degeneracy of the ground state by $\alpha,\beta, \ldots$, we get the effective Hamiltonian governing gauge dynamics:
\begin{equation}
H_{eff}^{\alpha\beta}=-\frac{g^2}{2}\mathcal{D}^{\alpha\gamma}_{ia}\mathcal{D}^{\gamma\beta}_{ia}+\delta^{\alpha\beta}\left(\frac{1}{g^2} V(M)+\epsilon_0(M)+\frac{g^2}{2}\Phi(M)\right)
\label{Heff}
\end{equation}
where $\mathcal{D}$ is the covariant derivative
\begin{equation}
\mathcal{D}^{\alpha\beta}_{ia}=\delta^{\alpha\beta}\partial_{ia}-i \mathcal{A}^{\alpha\beta}_{ia}
\end{equation}
and
\begin{equation}
\mathcal{A}^{\alpha\beta}_{ia}\equiv i\langle 0(M),\alpha|\partial_{ia}|0(M),\beta\rangle
\end{equation}
is the well-known Berry (or adiabatic) connection for the degenerate ground states $|0(M),\alpha\rangle$. The corresponding curvature is
\begin{equation}
\mathcal{F}_{ia,jb}^{\alpha \beta}\equiv \partial_{ia}\mathcal{A}_{jb}^{\alpha \beta}-\partial_{jb}\mathcal{A}_{ia}^{\alpha \beta}-i[\mathcal{A}_{ia},\mathcal{A}_{jb}]^{\alpha \beta}.
\label{adiacurv}
\end{equation}

Thus with the fermion in the ground state, gauge variables $M_{ia}$ feel an induced adiabatic connection. When the ground state is degenerate,
 this potential is non-Abelian.

The Hamiltonian (\ref{Heff}) has an additional effective scalar potential
\begin{equation}
\Phi= \sum_{l\neq 0} A_{ia}^{0l}A_{ia}^{l0}.
\label{scalarpotential}
\end{equation}
Note that the scalar potential has appeared simply because we have restricted to the ground state; if instead we had 
taken into account all the states in the spectrum of $H_f$ (equivalently, filled all available Fermi states), then we would have to work with the set of 
equations (\ref{Cn}), and no scalar potential would arise.

The scalar potential (\ref{scalarpotential}) can be  can be written in terms of the projector $P_0$ to the ground state \cite{kato,Zanardi}:
\begin{equation}
\Phi=\frac{1}{g_0}{\rm Tr}\left(P_0\partial_{ia}H_f\frac{Q_0}{(H-\epsilon_0)^2}\partial_{ia}H_f P_0\right)
\label{phiasP}
\end{equation}
where $g_0$ is the degeneracy and $Q_0=\mathbf{1}-P_0$.

Berry, in his article \cite{berry} on quantum adiabatic transport, relates the scalar potential to the quantum 
geometric tensor $G_{IJ}$. Given a quantum state $|n(x_I)\rangle$, where $x_{I}$ are the parameters (analogous 
to our gauge variables $M_{ia}$), one can define a gauge-invariant Hermitian tensor
\begin{equation}
G_{IJ}=\langle \partial_{I}n|Q_n|\partial_{J} n\rangle, \quad Q_n = {\bf 1} - |n(x_I)\rangle \langle n(x_I)|.
\end{equation}

Separating it into its real and imaginary parts
\begin{equation}
G_{IJ}=g_{IJ}+i\frac{F_{IJ}}{2}
\end{equation}
gives us $g_{IJ}$, the quantum metric tensor, which is positive-definite and symmetric. It provides a measure of distance between two quantum 
states separated in the parameter space. The imaginary part $F_{IJ}$ is just the adiabatic curvature corresponding to the parallel transport of 
state $|n(x_I)\rangle$. The scalar potential corresponding to $|n\rangle$ is  the trace of the metric $g_{IJ}$. 

The quantum geometric tensor has also been discussed in \cite{Zanardi} in relation to quantum fidelity.

It is useful to write the expression for $\Phi$ in terms of the eigenstates explicitly:
\begin{equation}
\Phi=\frac{1}{g_0}\sum_{n \neq 0}\sum_{ \alpha,\beta} \sum_{i,a} \frac{|\langle 0,\alpha|\partial_{ia}H_f|n,\beta\rangle|^2}{(\epsilon_n-\epsilon_0)^2}
\label{phiasState}
\end{equation}

\subsection{Effective Gauss' law}
What happens to the Gauss' law in the effective theory? After all, if the state $|\psi^E\rangle$ is physical, then it 
must be annihilated by the Gauss' law generators: 
\[G_a|\psi^E\rangle=0.\]

Following a procedure analogous to the one described in section $3$, we obtain the effective Gauss' law 
action on $\psi_E^m$
\begin{equation}
\mathcal{G}_a^{nm}\psi_E^{m}=0
\end{equation}
where
\begin{equation}
\mathcal{G}_a^{nm}= i \epsilon_{abc}M_{ib}\mathcal{D}_{ic}^{nm}-\frac{1}{2g^2}\langle n(M)|\lambda^\dagger(\mathbf{1}\otimes\tau_a)\lambda|m(M)\rangle
\label{Geff}
\end{equation}

The first term in the RHS of (\ref{Geff}) is the "covariantized" generator of gauge rotations in the gauge configuration space, while the second term is the generator of gauge rotations for the fermions.

Restricting to the fermion ground state and taking into account degeneracies, we have the effective Gauss' 
law generators
\begin{equation}
\mathcal{G}^{\alpha\beta}_a=i \epsilon_{abc}M_{ib}\mathcal{D}_{ic}^{\alpha\beta}-\frac{1}{2g^2}\langle 0(M),\alpha|\lambda^\dagger(\mathbf{1}\otimes \tau_a)\lambda|0(M),\beta\rangle.
\label{Gauss}
\end{equation}

The two terms of (\ref{Gauss}) can be combined to give
\begin{align*}
\mathcal{G}_a^{\alpha\beta}&=i\delta^{\alpha\beta}\epsilon_{abc}M_{ib}\partial_{ic}+i\epsilon_{abc}M_{ib}\langle 0(M),\alpha|\partial_{ic}|0(M),\beta\rangle-\frac{1}{2g^2}\langle 0(M),\alpha|\lambda^\dagger_{\alpha A}(\tau_a)_{AB}\lambda_{\alpha B}|0(M),\beta\rangle\\
&= i\delta^{\alpha \beta} \epsilon_{abc}M_{ib}\partial_{ic}+\langle 0(M),\alpha|G_a|0(M),\beta\rangle
\end{align*}

Since $h(M)$ is gauge-invariant, we can arrange for its eigenvectors to be annihilated by Gauss' law generators: $G_a|n(M)\rangle=0$ for any eigenstate $|n(M)\rangle$. Then the effective Gauss' law generator $\mathcal{G}_a^{\alpha\beta}$ is simply
\begin{equation}
\mathcal{G}_a^{\alpha\beta}=i\delta^{\alpha\beta}\epsilon_{abc}M_{ib}\partial_{ic}.
\end{equation}

A straightforward computation yields
\begin{equation}
[\mathcal{G}_a,\mathcal{G}_b]^{\alpha\beta}=-i\epsilon_{abc}\mathcal{G}^{\alpha\beta}.
\end{equation}

\section{Fermionic Spectrum}

Since $h(M)$ commutes with $\sum_{\alpha,A}\lambda^\dagger_{\alpha A}\lambda_{\alpha A}$, its eigenstates can be organized according to fixed fermion number. The most general ansatz for an $r$-fermion eigenstate is a linear combination of states with different spin and colour:
\begin{equation}
|n_{(r)}(M),r\rangle=f(M)^{n_{(r)}}_{\alpha_1A_1...\alpha_rA_r}\lambda^\dagger_{\alpha_1 A_1}...\lambda^\dagger_{\alpha_r A_r}|0\rangle
\label{rFermAnsatz}
\end{equation}
Because any two $\lambda^\dagger$'s anticommute, the $f(M)^{n_{(r)}}$ is antisymmetric under the exchange of any two pairs of indices $\alpha_i A_i$ and $\alpha_j A_j$.

Taking the scalar product of $(\ref{rFermAnsatz})$ with $\langle \lambda_{\alpha_1A_1}...\lambda_{\alpha_{r'}A_{r'}}|$ on both sides, we find that
\begin{equation}
C^{n_{(r)}}_{\alpha_1A_1...\alpha_{r'}A_{r'}}= \left\{ 
\begin{array}{c}
(g^2)^rf(M)^{n_{(r)}}_{\alpha_1A_1...\alpha_rA_r}, \quad  r=r' \\
0,\hspace{2.75cm} r\neq r'
\end{array} \right.
\end{equation}
%
%
Substituting (\ref{rFermAnsatz}) in (\ref{heqn}), we get
\begin{equation}
\sum_{i=1}^r g^2(H_f)_{\alpha_i A_i,\beta B} C^{n_{(r)}}_{\alpha_1 A_1...\hat{\alpha_i} \hat{A_i}\beta B...\alpha_rA_r}= \epsilon_{n_{(r)}} C^{n_{(r)}}_{\alpha_1A_1...\alpha_rA_r}
\label{fermioneqn}
\end{equation}
where the hat over an index denotes that index being excluded from the sum. 

Let us consider the single-particle sector first. The normalized eigenstates are of the form
\begin{equation}
|n\rangle =\frac{1}{g} C^n_{\alpha A}\lambda^\dagger_{\alpha A}|0\rangle
\end{equation}
where $C^n$ satisfies the eigenvalue equation
\begin{equation}
g^2(H_f)_{\alpha A,\beta B}C^n_{\beta B}=\epsilon_n C^n_{\alpha A}.
\end{equation}
Thus the $C^n$'s are the normalized eigenvectors of the $4\times 4$ matrix
\begin{equation}
g^2H_f(M)=-\mathbf{1}-\frac{1}{2}\sigma_i\otimes\tau_a M_{ia}
\end{equation}

Now since $h(M)$ commutes with the Gauss' law (\ref{gausslaw}), its eigenstates must be gauge invariant, implying that
\begin{equation}
G_a|n(M)\rangle=0
\end{equation}
or equivalently,
\begin{equation}
[G_a,C^n_{\alpha A}\lambda^\dagger_{\alpha A}]=0.
\end{equation}
This gives us the (gauge) transformation properties of the $C^n$'s:
\begin{align}
&\epsilon_{acd}[\Pi_{ic},C_{\alpha A}(M)]M_{id}\lambda_{\alpha A}^\dagger-(\tau_a)_{CD}C_{\alpha A}(M)\lambda_{\gamma C}^\dagger
\{\lambda_{\gamma D},\lambda_{\alpha A}^\dagger\}\\
&=-i\epsilon_{acd}\frac{dC_{\alpha A}(M)}{dM_{ic}}M_{id}\lambda_{\alpha A}^\dagger-(\tau_a)_{CD}C_{\alpha D}(M)\lambda_{\alpha C}^\dagger\\
&=i[-\epsilon_{acd}M_{id}\delta_{FG}\frac{dC_{\alpha G}(M)}{dM_{ic}}+i(\tau_a)_{FG}C_{\alpha G}(M)]\lambda^\dagger_{\alpha F}=0
\end{align}
implying that
\begin{eqnarray}
\label{difeq}
\epsilon_{abc}M_{ic}\frac{dC_{\alpha A}(M)}{dM_{ib}}=i(\tau_a)_{AB}C_{\alpha B}(M).
\end{eqnarray}
This can be explicitly seen by noting that under a gauge transformation, $M\rightarrow Mh^T$,$h\in SO(3)$,
\begin{align*}
g^2H_f(Mh^T)&= -\mathbf{1}-\frac{1}{2}\sigma_i\otimes\tau_a M_{ib}h_{ab}\\
&= -\mathbf{1}-\frac{1}{2}\sigma_i\otimes(g_{ab}\tau_a) M_{ib}h_{ab}\\
&= -\mathbf{1}-\frac{1}{2}\sigma_i\otimes(u(h)\tau_bu(h)^\dagger) M_{ib}h_{ab}\\
&=(\mathbf{1}\otimes u(h))( g^2H_f(M) )(\mathbf{1}\otimes u(h))^\dagger.
\end{align*}
So the eigenvectors of $H_f$ must transform as
\[ C(Mh^T)= (\mathbf{1}\otimes u(h))C(M)\]
i.e.,
\begin{equation}
C(Mh^T)_{\alpha A}=u(h)_{AB}C(M)_{\alpha B}.
\label{gaugeTrans2}
\end{equation}
Taking infinitesimal $h \simeq \mathbb{I}-iT_a\theta_a$ and noting that $M$ transforms in the adjoint while the $H_f$ transforms in the fundamental representation, we obtain (\ref{gaugeTrans2}) explicitly as the infinitesimal version of (\ref{difeq}).

Thus under a gauge transformation, $C(M)_{\alpha A}$ must transform in the fundamental (i.e. spin-$1/2$) 
representation of the gauge group. So the 
$C(M)$ defines a spinor field on the configuration space $\mathcal{C}$.

States with higher fermion numbers can be easily constructed out of the single fermion state. For example 
consider the 2-fermion state
\begin{equation}
|n_{(2)}\rangle= C^{n_{(2)}}_{\alpha_1 A_1 \alpha_2 A_2}\lambda^\dagger_{\alpha_1A_1}\lambda^\dagger_{\alpha_2A_2}|0\rangle, \quad C^{n_{(2)}}_{\alpha_1 A_1 \alpha_2 A_2}=-C^{n_{(2)}}_{\alpha_2 A_2 \alpha_1 A_1}.
\end{equation}
%
%
Then (\ref{fermioneqn}) reduces to
\[
g^2(H_f)_{\alpha_1 A_1,\beta B} C^{n_{(2)}}_{\beta B \alpha_2 A_2}+g^2(H_f)_{\alpha_2 A_2,\beta B} C^{n_{(2)}}_{\alpha_1 A_1\beta B}=\epsilon_{n_{(2)}}C^{n_{(2)}}_{\alpha_1A_1 \alpha_2A_2} \]
or equivalently
\begin{equation}
g^2(H_f\otimes\mathbf{1}+\mathbf{1}\otimes H_f)_{\alpha_1A_1\alpha_2A_2,\beta_1 B_1 \beta_2 B_2} C^{n_{(2)}}_{\beta_1 B_1 \beta_2 B_2}=\epsilon_{n_{(2)}}C^{n_{(2)}}_{\alpha_1A_1 \alpha_2A_2}
\end{equation}
where $(H_f\otimes\mathbf{1})_{\alpha_1 A_1 \alpha_2 A_2,\beta_1 B_1 \beta_2 B_2}=(H_f)_{\alpha_1 A_1,\beta_1 B_1}(\mathbf{1})_{\alpha_2 A_2,\beta_2B_2}$.

To make the notation simpler, let us represent the single-particle $C^n$ as a $6$-dimensional vector, and write
\[ g^2H_f|C^n\rangle=\epsilon_n|C^n\rangle.\]
%
%
%
With correct normalization, the two-fermion eigenstate can be written as
\begin{equation}
|n_{(2)}\rangle\equiv |n_1,n_2\rangle =\frac{1}{2g^2}(C^{n_1}_{\alpha_1 A_1}C^{n_2}_{\alpha_2 A_2}-C^{n_2}_{\alpha_1 A_1}C^{n_1}_{\alpha_2 A_2})\lambda^\dagger_{\alpha_1A_1}\lambda^\dagger_{\alpha_2A_2}
|0\rangle=\frac{1}{\sqrt{2}g^2}C^{n_{(2)}}_{\alpha_1 A_1 \alpha_2 A_2}\lambda^\dagger_{\alpha_1A_1}\lambda^\dagger_{\alpha_2A_2}
|0\rangle
\label{2fermionstate}
\end{equation}
with eigenvalue $\epsilon_{n_1}+\epsilon_{n_2}$.

Similarly, the $r$-fermion eigenstate is
\begin{equation}
|n_1,n_2,...n_r\rangle=\frac{1}{g^r} C^{n_1}_{[\alpha_1A_1}C^{n_2}_{\alpha_2 A_2}...C^{n_r}_{\alpha_r A_r]}\lambda^\dagger_{\alpha_1A_1}\lambda^\dagger_{\alpha_2A_2}...\lambda^\dagger_{\alpha_rA_r}
|0\rangle=\frac{1}{g^r\sqrt{r!}}C^{n_{(r)}}_{\alpha_1 A_1... \alpha_r A_r}\lambda^\dagger_{\alpha_1A_1}...\lambda^\dagger_{\alpha_rA_r}
|0\rangle
\end{equation}
with energy
\begin{equation}
h(M)|n_1,n_2...n_r\rangle=\left(\sum_{i=1}^r\epsilon_{n_i}\right)|n_1,n_2...n_r\rangle.
\end{equation}

Let us do a counting of the energy levels. Since $H_f$ is a $4\times 4$ matrix, there are $4$ single-particle energy levels. Higher fermion-number states can be obtained by putting fermions in each of these levels. So there are $^4C_r$ energy levels for an $r$-fermion state.

As it turns out, the matrix model for $SU(2)$ gauge theory coupled to a single Weyl fermion has a gauge 
anomaly \cite{pv1}, the very same as discovered by Witten \cite{Witten:1982fp}. In our quantum mechanical 
problem, the anomaly can be avoided by considering states with even number of fermions, or 
alternately, considering fermionic states with equal number of fermions of positive and negative chirality. It may seem that the one-fermion spectrum is of no physical importance, since one needs to work with even 
number of fermions for an anomaly-free theory. However, as we shall see in Section 7, it is essential 
to examine the one-fermion spectrum, not only because multi-particle spectra are constructed out of such 
states, but also because the one-fermion sector can capture topological features and functional 
analytic information about the "corners" of the gauge configuration space. These aspects become visible when we solve for the spectrum of the fermion Hamiltonian $H_f(M)$ in the background of $M_{ia}$.

\subsection{Characteristic Polynomial of $H_f$}

 The $-\mathbf{1}$ in $g^2 H_f$ only adds the same overall constant to all the energy levels, and can be ignored by a simple redefinition of the zero of the energy. We define
 \[ (H'_f)\equiv g^2 H_f+\mathbf{1}\]
 and henceforth drop the prime, thus working with
\begin{eqnarray}
\label{hamilFund}
(H_f)_{\alpha A,\beta B}=
-\frac{1}{2}(\tau_c)_{AB}(\sigma^i)_{\alpha\beta}M_{ic}.
\end{eqnarray}
The characteristic equation of $H_f$ is
\begin{eqnarray}
\hat{\lambda}^4-\frac{\hat{\lambda}^2}{2}\textrm{Tr}(M^TM)+\hat{\lambda}\det M
+\frac{1}{16}\left[2\textrm{Tr}(M^TM)^2-(\textrm{Tr}(M^TM))^2\right]=0.
\end{eqnarray}
Rescaling
\begin{eqnarray}
x=\frac{\hat{\lambda}}{\left(\frac{1}{3}\textrm{Tr}(M^TM)\right)^{1/2}} \equiv \frac{\hat{\lambda}}{{\mathbf g}_2},
\end{eqnarray}
we obtain the characteristic equation in terms of scale-invariant dimensionless quantities $\mathbf{g}_3$ and $\mathbf{g}_4$
\begin{eqnarray}
x^{4}-\frac{3}{2}x^{2}-{\mathbf g}_3 x+ {\mathbf g}_4=0,
\label{cheq}
\end{eqnarray}
with
\begin{eqnarray}
{\mathbf g}_3\equiv \frac{\det M}{\left(\frac{1}{3}\textrm{Tr}(M^TM)\right)^{3/2}}, \quad\quad 
{\mathbf g}_4 \equiv \frac{1}{16}\left[\frac{2\textrm{Tr}(M^TM)^2}{\left(\frac{1}{3}\textrm{Tr}(M^TM)\right)^2}-9\right].
\end{eqnarray}
The variables $\mathbf{g}_2,\mathbf{g}_3$ and $\mathbf{g}_4$ are gauge- and rotationally invariant independent quantities. They may be thought of 
as 3 of the coordinates on the gauge configuration space ${\cal C}$, the other 3 being physical rotations.

Since $H_f$ is Hermitian, its eigenvalues, and hence roots of  (\ref{cheq}) must be real. The nature of the roots can be determined by studying the discriminant $\Delta$ of the quartic polynomial, given by
\begin{eqnarray}
\Delta=\frac{1}{2}(27{\mathbf g}_3^2-54{\mathbf g}_3^4+162{\mathbf g}_4-432{\mathbf g}_3^2{\mathbf g}_4-576{\mathbf g}_4^2+512{\mathbf g}_4^3),
\end{eqnarray}
and four other quantities \cite{quartic}
\begin{eqnarray}
P=-12, \quad\quad Q=-8{\mathbf g}_3, \quad\quad \Delta_0=\frac{9}{4}+12{\mathbf g}_4, \quad\quad D=4(16{\mathbf g}_4-9).
\end{eqnarray}
For all the roots to be real, we must have $P<0$ and
\begin{equation}
\Delta \geq 0 \quad {\rm and} \quad D\leq 0.
\end{equation}
This is a simple reformulation of Sylvester's theorem on the real roots of a real polynomial (see for eg \cite{prasolov}). As a by-product, we have thus found two inequalities that ${\mathbf g}_3$ and ${\mathbf g}_4$, and hence {\it all} real $3\times3$ matrices must obey. 

The region $\Delta \geq0$ in the ${\mathbf g}_3$-${\mathbf g}_4$ space is shown in Fig. 1.

\begin{figure}[hbtp]
\centering
\includegraphics[scale=0.4]{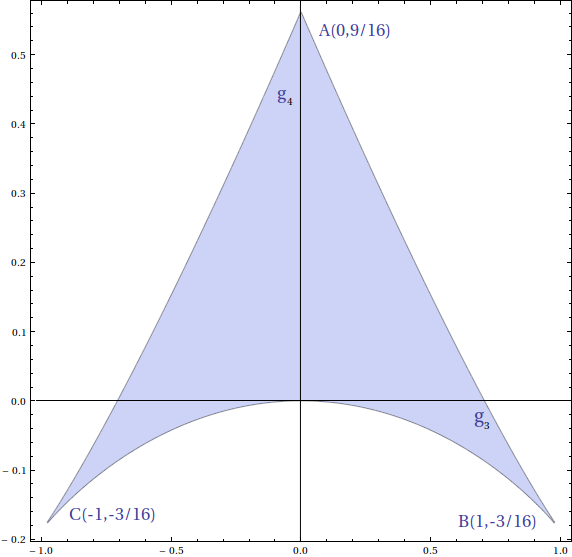}  
\caption{\texttt{Allowed region in $\mathbf{g}_3$-$\mathbf{g}_4$ space}}
\end{figure}
The matrix $M$ permits a singular value decomposition (SVD) as $M=RAS^T$ where $R,S\in SO(3)$
and $A=diag(a_1,a_2,a_3)$ with $a_1\geq a_2\geq |a_3| \geq 0$. Using the explicit expressions of ${\mathbf g}_3$ and ${\mathbf g}_4$, we find that
\begin{eqnarray}
\Delta=\frac{(a_1^2-a_2^2)^2(a_1^2-a_3^2)^2(a_2^2-a_3^2)^2}{[\frac{1}{3}(a_1^2+a_2^2+a_3^2)]^6}.
\end{eqnarray}
and
\begin{eqnarray}
D=-144\frac{a_1^2a_2^2+a_1^2a_3^2+a_2^2a_3^2}{(a_1^2+a_2^2+a_3^2)^2}.
\end{eqnarray}
That is, $\Delta\geq 0$ and $D\leq 0$ identically.

The edges and corners of Fig. 1 correspond to $\Delta=0=D$, and are places where two or more of the singular values coincide. At the edge $AB$ we have $a_2=a_3$, at $AC$ we have $a_2=-a_3$ and at $BC$ we have $a_1=a_2$. The point $A$ corresponds to $a_2=a_3=0$, the point $B$ to $a_1=a_2=a_3$ and $C$ to $a_1=a_2=-a_3$.

We can also express the curves for the edges in terms of variables $\mathbf{g}_3$ and $\mathbf{g}_4$:
\begin{eqnarray}
36\mathbf{g}_3^2 &=& -144 \mathbf{g}_4- \sqrt{3(16 \mathbf{g}_4+3)^3}+9 \quad {\rm for} \,\,BC, \label{curveBC}\\
6\mathbf{g}_3 &=& +\sqrt{-144 \mathbf{g}_4+\sqrt{3(16 \mathbf{g}_4+3)^3}+9} \quad {\rm for} \,\,AB, \label{curveAB}\\
6\mathbf{g}_3 &=& -\sqrt{-144 \mathbf{g}_4+ \sqrt{3(16 \mathbf{g}_4+3)^3}+9} \quad {\rm for} \,\,AC. \label{curveAC}
\end{eqnarray}

Since the spectrum of $H_f$ is gauge invariant, the energy eigenvalues can be expressed in terms of gauge-invariant functions of $M$. In fact
\begin{equation}
\epsilon_n(M)=\epsilon_n({\mathbf g}_2,{\mathbf g}_3,{\mathbf g}_4)={\mathbf g}_2 
x_n({\mathbf g}_3,{\mathbf g}_4)
\end{equation}
We therefore expect the spectrum to carry information about the edges and corners of Fig. 1. It turns out that these are places where the degeneracies of fermionic energy levels change. 

\subsection{Eigenfunctions of $H_f$}

For the $SU(2)$ theory, we can obtain the spectrum and eigenfunctions of $H_f$ in a very simple form in terms of the singular values of $M_{ia}$. Writing $M=RAS^T$, we see that $M$ is related to $A$ by a rotation $R$ and a gauge transformation $S$.

Under a combined rotation and gauge transformation, $H_f$ transforms as
\begin{equation}
H_f(RMS^T)=(r(R)\otimes s(S)) H_f(M) (r(R)\otimes s(S))^\dagger.
\label{Htrans}
\end{equation}
Here $r(R)$ and $s(S)$ are spin-$1/2$ representations of $R$ and $S$ respectively. Thus $H_f(M)$ and $H(A)$
are unitarily related, their spectra are the same, and the eigenfunction
\begin{equation}
C(M)=(r(R)\otimes s(S))C(A)
\label{Ctrans}
\end{equation}

So we can work with the SVD of $M$ and take the Hamiltonian to be
\begin{equation}
H(A)=-\frac{1}{2}\sum_{i=1}^3 a_i\sigma_i\otimes \tau_i .
\end{equation}
The one-fermion eigenfunctions and eigenvalues are
\begin{subequations}
\begin{align}
&|C^1\rangle= \frac{1}{\sqrt{2}}\Big(\left|\textstyle{\frac{1}{2}}\right\rangle \left|-\textstyle{\frac{1}{2}}\right\rangle
+ \left|-\textstyle{\frac{1}{2}}\right\rangle \left|\textstyle{\frac{1}{2}}\right\rangle\Big)\\
&|C^2\rangle=\frac{1}{\sqrt{2}}\Big(\left|\textstyle{\frac{1}{2}}\right\rangle \left|\textstyle{\frac{1}{2}}\right\rangle
+ \left|-\textstyle{\frac{1}{2}}\right\rangle \left|\textstyle{-\frac{1}{2}}\right\rangle\Big)\\
&|C^3\rangle=\frac{1}{\sqrt{2}}\Big(\left|\textstyle{\frac{1}{2}}\right\rangle \left|\textstyle{\frac{1}{2}}\right\rangle
- \left|-\textstyle{\frac{1}{2}}\right\rangle \left|-\textstyle{\frac{1}{2}}\right\rangle\Big)\\
&|C^4\rangle=\frac{1}{\sqrt{2}}\Big(\left|\textstyle{\frac{1}{2}}\right\rangle \left|-\textstyle{\frac{1}{2}}\right\rangle
- \left|-\textstyle{\frac{1}{2}}\right\rangle \left|\textstyle{\frac{1}{2}}\right\rangle\Big)
\end{align}
\label{1pspec}
\end{subequations}

\begin{subequations}
\begin{align}
&\epsilon_1=\frac{1}{2}(-a_1-a_2+a_3)\\
&\epsilon_2=\frac{1}{2}(-a_1+a_2-a_3)\\
&\epsilon_3=\frac{1}{2}(a_1-a_2-a_3)\\
&\epsilon_4=\frac{1}{2}(a_1+a_2+a_3).
\end{align}
\label{1penergies}
\end{subequations}

Here our conventions are
\[\sigma_{3}\left|\pm \textstyle{\frac{1}{2}}\right\rangle=\pm \left|\pm \textstyle{\frac{1}{2}}\right\rangle;\quad 
\tau_{3}\left|\pm \textstyle{\frac{1}{2}}\right\rangle=\pm \left|\pm \textstyle{\frac{1}{2}}\right\rangle.\]

The energies can be written in terms of matrix invariants $\mathbf{g}_2, \mathbf{g}_3$ and $\mathbf{g}_4$ since 
$\epsilon_i=\mathbf{g}_2 x_i(\mathbf{g}_3,\mathbf{g}_4)$, and $x_i$ are the roots of the quartic polynomial 
$(\ref{cheq})$. This form is useful because it is written in terms of manifestly gauge (and rotation) 
invariant quantities.

The rescaled energy levels $x_i$ are plotted against the invariants $\mathbf{g}_3$ and $\mathbf{g}_4$ in Fig. 2. The $x_i$'s can also be 
determined explicitly in terms of $\mathbf{g}_3$ and $\mathbf{g}_4$, but we will not present these expressions here.  As functions of $\mathbf{g}_3$ and $\mathbf{g}_4$, they are non-analytic at the edges and corners of Fig. 1.
\begin{figure}[hbtp]
\centering
\includegraphics[scale=.4]{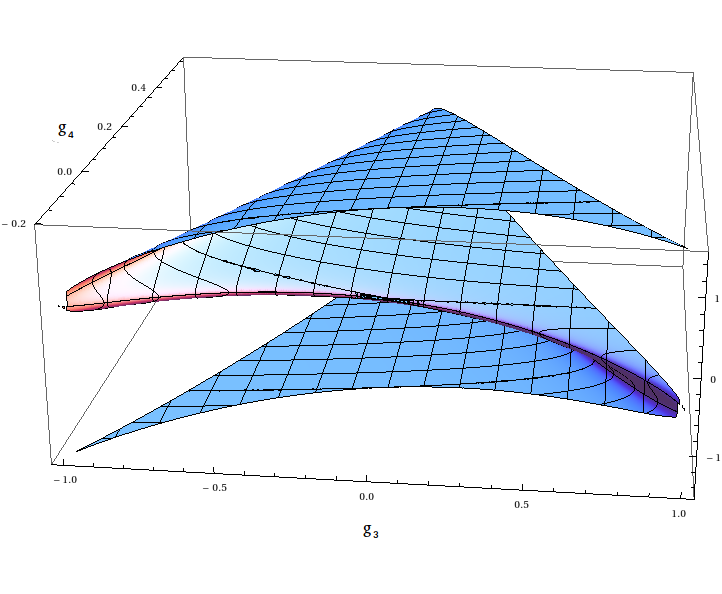}
\caption{\texttt{Plot of $x_i$ against $\mathbf{g}_3,\mathbf{g}_4$}}
\end{figure}

\begin{figure}[hbtp]
\centering
\includegraphics[scale=.4]{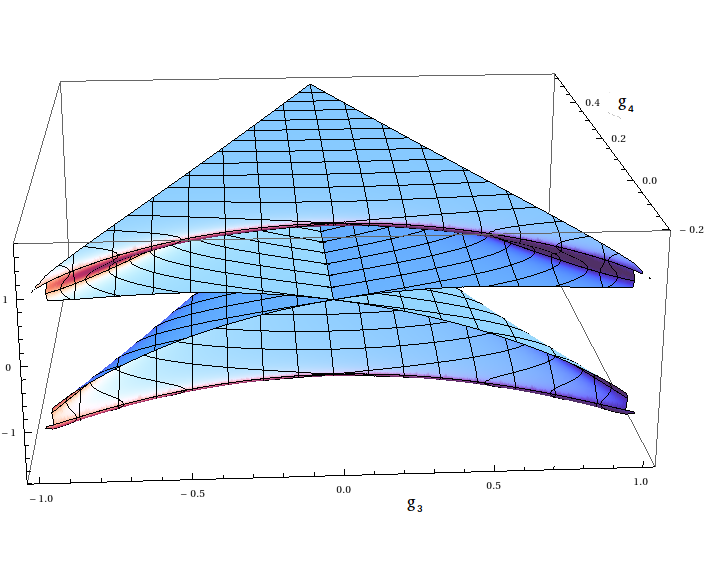}
\caption{\texttt{Plot of 2-fermion energies against $\mathbf{g}_3,\mathbf{g}_4$}}
\end{figure}
Note that since Tr $H_f=0$, we have $\sum_{i=1}^4 \epsilon_i=0$. So filling all the energy levels gives back total energy zero, just like the vacuum; filling only $3$ levels gives negative of the energy of the unfilled level, and so on. This gives rise to fermion-hole correspondence:
the $3$-fermion spectrum is analogous to the $1$-hole picture, and the energy levels are just the negatives of the one-fermion energy levels; the $4$-fermion spectrum is like the hole vacuum. The two-fermion spectrum is the dual of itself and hence it is symmetric about the zero of the energy.
The two-fermion energies are
\begin{subequations}
\begin{align}
&\epsilon^{(2)}_1=-a_1\\
&\epsilon^{(2)}_2=-a_2\\
&\epsilon^{(2)}_3=-a_3\\
&\epsilon^{(2)}_4=a_3\\
&\epsilon^{(2)}_5=a_2\\
&\epsilon^{(2)}_6=a_1
\end{align}
\end{subequations}
Knowing the energies explicitly allows us to write down the characteristic polynomial for the 2-fermion Hamiltonian
\begin{equation}
H_f^{(2)} = H_f\otimes\mathbf{1}+\mathbf{1}\otimes H_f
\end{equation}
quite easily. Rather than write it as a function fo the $a_i$'s we will present it here in terms of $\mathbf{g}_2,\mathbf{g}_3$ and $\mathbf{g}_4$, 
which will turn out more useful form:
\begin{equation}
\mathbf{g}_2^6(y^6-3y^4+4y^2(9/16-\mathbf{g}_4)-\mathbf{g}_3^2)=0.
\label{2fchar}
\end{equation}

\subsection{Enhanced Symmetry at the Corners}

From $(\ref{Htrans})$, $H_f$ transforms under the group $SU(2)_{spin}\times SU(2)_{color}$. What are the symmetries of $H_f$? The form of 
$H_f(M=A)$ suggests that there is a coupling between the spin and colour, and as we will see below, the eigenstates of the Hamiltonian arrange themselves in multiplets that transform under a total angular momentum of a spin-colour coupling.

Let us define a "total angular momentum" operator\footnote{In \cite{Schafer:2000tw}, this is called "grand angular momentum".} as
\begin{equation}
J_i=\frac{1}{2}(\sigma_i\otimes\mathbf{1}+\mathbf{1}\otimes \tau_i).
\label{totalJ}
\end{equation}
It can be seen that
\begin{equation}
[J^2,H_f]=0
\end{equation}
but, in general, $[J_i,H]\neq 0$. In fact:
\begin{equation}
[J_i,H_f]=\frac{i}{2}\sum_{j,k}\epsilon_{ijk}(a_j-a_k)\sigma_j\otimes\tau_k.
\label{commJH}
\end{equation}
So in general when the $a_i$'s are all different, there is no Lie algebra element of $SU(2)_{spin}\times SU(2)_{color}$ that commutes with $H_f$. 
At the corners where two or more singular values coincide, there are new symmetries of $H_f$.

Eq. $(\ref{totalJ})$ suggests that the energy eigenstates transform under irreps of $SU(2)$ obtained by adding two spin-$1/2$ representations.  Labelling eigenstates of $J^2$ and $J_3$ as $(l,m)$, we can easily see that
\begin{subequations}
\begin{align}
&|1\rangle= |1,0\rangle\\
&|2\rangle= \frac{1}{\sqrt{2}}\Big(|1,1\rangle+|1,-1\rangle\Big)\\
&|3\rangle= \frac{1}{\sqrt{2}}\Big(|1,1\rangle-|1,-1\rangle\Big)\\
&|4\rangle= |0,0\rangle
\end{align}
\end{subequations}

So $|1\rangle,|2\rangle,|3\rangle$ form the triplet (spin-1) and $|4\rangle$ the singlet (spin-0).

At the different edges and corners, there are enhanced symmetries of $H_f$:

\begin{enumerate}

\item $a_1=a_2=a > a_3 \geq 0$\\
  From $(\ref{commJH})$, one can see that $[J_3, H_f]=0$, so a combined rotation and gauge transformation around the {\it third} axis leaves $H_f$ invariant. Here, $\epsilon_2=\epsilon_3=-\frac{a_3}{2}$, so any linear combination of $|2\rangle$ and $|3\rangle$ are eigenstates of $H_f$ with same eigenvalue, and in particular $|1,1\rangle$ and $|1,-1\rangle$. Thus energy eigenstates are of the form $|j,m\rangle$.

\item $a_1>a_2=a_3 =a\geq 0$\\
Here, $[J_1, H_f]=0$, so a combined rotation and gauge transformation around the {\it first} axis leaves the 
Hamiltonian invariant. The lowest energy level is degenerate: $\epsilon_1=\epsilon_2=-\frac{a_1}{2}$. 
The states $|1\rangle$ and $|2\rangle$ can be combined to form eigenstates of $J_1$. So energy eigenstates are 
of the form $|j,m_x\rangle$, where the subscript $x$ denotes that the spin projection is in the first direction.
  
\item $a_1=a_2=a_3=a \neq 0$\\
In this corner, $[J_i,H_f]=0 \quad \forall i$, so combined rotation and gauge transformation about any axis leaves $H_f$ invariant. This is the maximally symmetric case. Here $\epsilon_1=\epsilon_2=\epsilon_3=-\frac{a}{2}$, and $\epsilon_4=\frac{3a}{2}$. The three degenerate ground states form a triplet under grand angular momentum 
$\vec{J}$, and the highest energy state a singlet.

\end{enumerate}

\section{Adiabatic Connection at different corners}

If the ground state is degenerate (the degeneracy labels being $\alpha,\beta, \ldots$), the adiabatic connection, in general non-Abelian is

\begin{equation}
\mathcal{A}^{\alpha\beta}=\mathcal{A}^{\alpha\beta}_{ia}dM_{ia}=i\langle0(M),\alpha|d|0(M),\beta\rangle.
\end{equation}
For the single-fermion states this gives
\begin{equation}
\mathcal{A}^{\alpha\beta}=i\bar{C}^{0,\alpha}(M)dC^{0,\alpha}(M)\equiv i\langle C(M)^{0,\alpha}|d|C(M)^{0,\beta}\rangle.
\end{equation}
Here, we have omitted the spin and colour index.

In SVD, this becomes
\begin{equation}
\mathcal{A}^{\alpha\beta}=i\langle C(A)^{0,\alpha}|\partial_{a_i}|C(A)^{0,\beta}\rangle da_i+i\langle C(A)^{0,\alpha}|(r(R)^\dagger dr(R))\otimes \mathbf{1}+\mathbf{1}\otimes (s(S)^\dagger ds(S))|C(A)^{0,\beta}\rangle.
\end{equation}

Let us define $\Omega_r\equiv r^\dagger dr$ and $\Omega_s\equiv s^\dagger ds$, left-invariant Maurer-Cartan forms on $SU(2)_{spin}$ and $SU(2)_{colour}$ respectively, in the fundamental representation. They are Lie algebra elements,and can be expanded as
\begin{equation}
\Omega_r=-i\omega_i^r\frac{\sigma_i}{2};\quad\Omega_s=-i\omega_a^s\frac{\tau_a}{2} 
\end{equation}
where $\omega_i^r$ and $\omega_a^s$ are real-valued one-forms (since the left-invariant form is 
anti-Hermitian, the factor of $-i$ ensures that the $\omega$ are real), whose exact form depends on 
the parametrizations of $r$ and $s$. They satisfy the equations
\begin{equation}
d\Omega_{r,s}+\Omega_{r,s}\wedge\Omega_{r,s}=0
\end{equation}
implying
\begin{equation}
d\omega_i+\frac{1}{2}\epsilon_{ijk}\omega_j\wedge\omega_k=0.
\end{equation}
From $(\ref{1pspec})$, it is clear that the $C(A)$'s are independent of the $a_i$'s. So the adiabatic connection is
\begin{equation}
\mathcal{A}^{\alpha\beta}=\frac{1}{2}\left(\langle C(A)^{0,\alpha}|(\omega_i^r\sigma_i)\otimes \mathbf{1}+\mathbf{1}\otimes (\omega_a^s\tau_a)|C(A)^{0,\beta}\rangle\right)
\end{equation}
and associated adiabatic curvature is
\begin{equation}
\mathcal{F}^{\alpha\beta}=(d\mathcal{A}+\mathcal{A}\wedge \mathcal{A})^{\alpha\beta}.
\end{equation}
 In the bulk (i.e. in the region $\Delta>0$), the ground state is non-degenerate:
\[|C(A)^0\rangle=\frac{1}{\sqrt{2}}\Big(\left|\textstyle{\frac{1}{2}}\right\rangle \left|-\textstyle{\frac{1}{2}}\right\rangle
+ \left|-\textstyle{\frac{1}{2}}\right\rangle \left|\textstyle{\frac{1}{2}}\right\rangle\Big).\]

The adiabatic connection for this state works out to be
\begin{align*}
\mathcal{A} &=\frac{1}{2}\cdot\frac{1}{2}\Big(\left\langle \textstyle{\frac{1}{2}}\right| 
\left\langle -\textstyle{\frac{1}{2}}\right| + \left\langle -\textstyle{\frac{1}{2}}\right| 
\left\langle \textstyle{\frac{1}{2}}\right|\Big)\Big( \omega_i^r\sigma_i\otimes \mathbf{1} + 
\mathbf{1}\otimes \omega_a^s\tau_a\Big) \Big(\left|\textstyle{\frac{1}{2}}\right\rangle 
\left|-\textstyle{\frac{1}{2}}\right\rangle + \left|-\textstyle{\frac{1}{2}}\right\rangle 
\left|\textstyle{\frac{1}{2}}\right\rangle\Big)\\
&= \frac{1}{4}\Big[\omega_i^r \Big(\left\langle \textstyle{\frac{1}{2}}\right| \sigma^i 
\left|\textstyle{\frac{1}{2}}\right\rangle + \left\langle -\textstyle{\frac{1}{2}}\right| \sigma^i 
\left|-\textstyle{\frac{1}{2}}\right\rangle\Big) + 
\omega_a^s \Big(\left\langle \textstyle{\frac{1}{2}}\right| \tau^a 
\left|\textstyle{\frac{1}{2}}\right\rangle + \left\langle -\textstyle{\frac{1}{2}}\right| \tau^a 
\left|-\textstyle{\frac{1}{2}}\right\rangle\Big)\Big]\\
&=\frac{1}{4}\left(\omega_i^r {\rm Tr}\,\sigma^i + \omega_a^s {\rm Tr}\,\tau^a \right)=0
\end{align*}
Thus in the bulk, the adiabatic connection vanishes, and so does the curvature:
\[\mathcal{F}_{bulk}=0.\]
At the corner $A$, and also along the edge $AB$, we have $a_1\geq a_2=a_3 \geq 0$. The two degenerate ground states are 
$|C^1\rangle$ and $|C^2\rangle$ of (\ref{1pspec}) and hence $\mathcal{A}$ is a non-abelian $U(2)$ matrix.
By a similar calculation as above, we find that the diagonal elements $\mathcal{A}_{11}$ and $\mathcal{A}_{22}$ are again 0, since 
they are proportional to traces of Pauli matrices. The off-diagonal element is, however, non-zero:
\begin{align*}
\mathcal{A}_{12} &=\frac{1}{2}\cdot\frac{1}{2}\Big[\left\langle \textstyle{\frac{1}{2}}\right| 
\left\langle -\textstyle{\frac{1}{2}}\right| + \left\langle -\textstyle{\frac{1}{2}}\right| 
\left\langle \textstyle{\frac{1}{2}}\right|\Big)\Big( \omega_i^r\sigma_i\otimes \mathbf{1} + 
\mathbf{1}\otimes \omega_a^s\tau_a\Big) \Big(\left|\textstyle{\frac{1}{2}}\right\rangle 
\left|\textstyle{\frac{1}{2}}\right\rangle + \left|-\textstyle{\frac{1}{2}}\right\rangle 
\left|-\textstyle{\frac{1}{2}}\right\rangle\Big] \\
&= \frac{1}{4}\Big[\omega_i^r \Big(\left\langle \textstyle{\frac{1}{2}}\right| \sigma^i 
\left|-\textstyle{\frac{1}{2}}\right\rangle + \left\langle -\textstyle{\frac{1}{2}}\right| \sigma^i 
\left|\textstyle{\frac{1}{2}}\right\rangle\Big) + 
\omega_a^s \Big(\left\langle -\textstyle{\frac{1}{2}}\right| \tau^a 
\left|\textstyle{\frac{1}{2}}\right\rangle + \left\langle \textstyle{\frac{1}{2}}\right| \tau^a 
\left|-\textstyle{\frac{1}{2}}\right\rangle\Big)\Big]\\
&=\frac{1}{2}(\omega_1^r+\omega_1^s)=\mathcal{A}_{12}^*
\end{align*}
%
The corresponding curvature is \[\mathcal{F}_{21}=\frac{1}{2} (d\omega_1^r +d\omega_1^s) = \mathcal{F}_{12}^*, \quad \mathcal{F}_{11}=\mathcal{F}_{22}=0.\]
So $\mathcal{F}\neq 0$ in the corner $A$ and the edge $AB$.

At the corner $B$, where all singular values coincide, $M$ is of the form $M=a (RS^T)=a G, G\in SO(3)$. 
We can work with only one matrix, say $R$ (equivalently set $S=1$; then $G=R$). Then $\mathcal{A}$ becomes
\begin{equation}
\frac{1}{2}\left(\langle C(A)^{0,\alpha}|(\omega_i^r\sigma_i)\otimes \mathbf{1}|C(A)^{0,\beta}\rangle\right).
\end{equation}
Here the ground state is triply degenerate, the states being $|C^1\rangle,|C^2\rangle,|C^3\rangle$ of (\ref{1pspec}). A short calculation yields $\mathcal{A}$ 
as a $3\times 3$ matrix
\begin{equation}
\mathcal{A}_{corner}=\omega_i^rT_i
\end{equation}
where
\begin{equation}
T_1=\left(\begin{array}{ccc}
0 & 1 & 0 \\ 
1 & 0 & 0 \\ 
0 & 0 & 0
\end{array}\right), \quad T_2=\left(\begin{array}{ccc}
0 & 0 & i \\ 
0 & 0 & 0 \\ 
-i & 0 & 0
\end{array}\right), \quad T_3=\left(\begin{array}{ccc}
0 & 0 & 0 \\ 
0 & 0 & 1 \\ 
0 & 1 & 0
\end{array}\right).
\end{equation}
Since \[[T_i,T_j]=i\epsilon_{ijk}T_k,\] the $T_i$ form the $3$-dimensional UIR of $SU(2)$. The $\mathcal{A}$ looks like the Maurer-Cartan form $\Omega_r$, but actually $\mathcal{A}=i\Omega$, which does not satisfy the structure equation
\begin{equation}
d(i\Omega)+(i\Omega)\wedge(i\Omega)\neq 0
\end{equation}
So in this corner too, we obtain a non-zero $\mathcal{F}$.

Thus in the bulk, the adiabatic curvature is zero. It is also straightforward to see that:
\begin{align*}
\frac{1}{2g^2}\langle C(A)^0|(\mathbf{1}\otimes \tau_a)|C(A)^0\rangle &= \frac{1}{2g^2}\frac{1}{2}\left(\langle \textstyle{\frac{1}{2}}| \tau_a|\textstyle{\frac{1}{2}}\rangle+\langle -\textstyle{\frac{1}{2}}| \tau_a|-\textstyle{\frac{1}{2}}\rangle \right)\\
&=\frac{1}{4g^2}({\rm Tr }\ \tau_a)=0
\end{align*}

The above discussion was for the one-fermion case. For the two-fermion case, it can be verified that
\begin{equation}
\langle n_1,n_2|d|n_3,n_4\rangle =\langle C^{n_1,n_2}|(d\otimes\mathbf{1}+\mathbf{1}\otimes d)|C^{n_3,n_4}\rangle
\end{equation}
where the 2-fermion eigenstate $|n_3,n_4\rangle$ is given in (\ref{2fermionstate}).
Because of the $\mathbf{1}$, the above matrix element is zero unless there is an overlap of at least one 1-fermion state between the two 2-fermion states. This allows us to compute the adiabatic connection for the 2-fermion states rather easily.

In the bulk, the ground state is non-degenerate, and is obtained by putting fermions in the first and the second energy levels. So
\begin{equation}
|C^0_{(2)}\rangle=\frac{1}{\sqrt{2}}\Big(|C^1\rangle|C^2\rangle-|C^2\rangle|C^1\rangle\Big).
\end{equation}
The adiabatic connection is
\begin{align*}
\mathcal{A}^{(2-fermion)}_{bulk}&=i\langle C^0_{(2)}| (d\otimes \mathbf{1}+\mathbf{1}\otimes d)|C^0_{(2)}\rangle\\
&=\frac{i}{2}\Big(2\langle C^1|d|C^1\rangle+2\langle C^2|d|C^2\rangle\Big)\\
&=\mathcal{A}^1_{bulk}+\mathcal{A}^2_{bulk}\\
&=0
\end{align*}
As expected, $\mathcal{A}_{bulk}=0$ in the 2-fermion sector as well.

The ground state degeneracy changes along the edge $BC$, where the singular values are $a_1=a_2> a_3\geq 0$. The two degenerate ground states are now
\[
|C^{0,1}_{(2)}\rangle=\frac{1}{\sqrt{2}}\Big(|C^1\rangle|C^2\rangle-|C^2\rangle|C^1\rangle\Big)
\]
and
\[
|C^{0,2}_{(2)}\rangle=\frac{1}{\sqrt{2}}\Big(|C^1\rangle|C^3\rangle-|C^3\rangle|C^1\rangle\Big).
\]
The adiabatic connection is again a $U(2)$ matrix. Its diagonal elements turn out to be $0$:
\begin{align*}
&\mathcal{A}^{2-fermion}_{11}=\mathcal{A}^1+\mathcal{A}^2=0,\\
&\mathcal{A}^{2-fermion}_{22}=\mathcal{A}^1+\mathcal{A}^3=0.
\end{align*}

The off-diagonal elements survive, giving
\begin{align*}
\mathcal{A}^{2-fermion}_{12}&=i\langle C^{0,1}_{(2)}| (d\otimes \mathbf{1}+\mathbf{1}\otimes d)|C^{0,2}_{(2)}\rangle\\
&=\frac{1}{2}(\omega_3^r+\omega_3^s)
\end{align*}
and $\mathcal{A}_{21}=\mathcal{A}_{12}^*=\mathcal{A}_{12}$. Thus, $\mathcal{A}^{2-fermion}_{edge}$ is 
non-zero, as is the adiabatic curvature.

Finally, at the corner $B$, where all singular values coincide, and degeneracy becomes $3$, we
find that
\begin{equation}
\mathcal{A}^{2-fermion}_{corner}=\omega_i^r T'_i
\end{equation}
where
\begin{equation}
T'_1=\left(\begin{array}{ccc}
0 & 0 & 0 \\ 
0 & 0 & 1 \\ 
0 & 1 & 0
\end{array}\right);\quad T_2'=\left(\begin{array}{ccc}
0 & 0 & -i \\ 
0 & 0 & 0 \\ 
i & 0 & 0
\end{array}\right);\quad T_3'=\left(\begin{array}{ccc}
0 & 1 & 0 \\ 
1 & 0 & 0 \\ 
0 & 0 & 0
\end{array}\right) 
\end{equation}
Again, the $T_i'$'s obey $[T_i',T_j']=i\epsilon_{ijk}T_k'$.

\section{Effective Scalar Potential}
The scalar potential (\ref{phiasP}) depends on the eigenstate of the fermion(s) via its dependence on the 
projector $P$. We are interested in the situation when the fermions are in the ground state of $H_f$. The projector $P_0$ to the ground state, however, changes rank depending on whether the external gauge variables correspond to a point in the 
bulk (of Fig. 1), an edge, or a corner. For each of these cases, (\ref{phiasP}) can be 
computed separately, and we will call them $\Phi_{bulk}, \Phi_{edge}$ and $\Phi_{corner}$. We will show that whenever the degeneracy of the ground state changes, the scalar potential shows a discontinuous behaviour.  More precisely, $\Phi_{bulk}$ diverges as we approach the edge $AB$, but $\Phi_{edge}$ is 
well-defined along almost the entire edge $AB$. Both $\Phi_{bulk}$ and $\Phi_{edge}$ diverge as we approach the point $B$, but $\Phi_{corner}$ is well-defined at $B$. This peculiar behaviour of $\Phi$ is the key reason for the emergence of superselection sectors in the Hilbert space for gauge dynamics, as we shall show in Section 7.


Although a 1-fermion state in the $SU(2)$ matrix model is anomalous, the computation of the scalar potential 
for this sector is both instructive and useful. For one, we can get an intuitive understanding of the relationship between the singularity structure of $\Phi$ and the change in the degeneracies of the ground state. Secondly, many of the formulas we derive for this case are useful when computing the scalar potential for the 2-fermion state. 

For a 1-fermion state with $M=RAS^T$, the effective scalar potential in the bulk  is
\begin{equation}
\Phi_{bulk}=\frac{2}{(a_2-a_3)^2}+\frac{2}{(a_1-a_3)^2}+\frac{2}{(a_1+a_2)^2}.
\end{equation}
The calculation is fairly straightforward if one uses (\ref{phiasState}) and the explicit expressions (\ref{1penergies}) and (\ref{1pspec}). 

$\Phi$ can be written in terms of the dimensionless quantities $\mathbf{g}_2,\mathbf{g}_3$ and $\mathbf{g}_4$ and the lowest root $x_1= x_1(\mathbf{g}_3,\mathbf{g}_4)$ of (\ref{cheq}) as
\begin{equation}
\Phi_{bulk}=\frac{1}{\mathbf{g}_2^2}\frac{8 \mathbf{g}_3 x_1+4(x_1^2 + 3/4)^2}{(-\mathbf{g}_3+4 x_1 (x_1^2 - 3/4))^2}.
\label{1fphibulk}
\end{equation}

To understand the singularity structure of $\Phi_{bulk}$, let us look at Fig. 4, which is a schematic diagram for the (rescaled) 1-fermion energy levels at different regions of 
the configuration space ($\mathbf{g}_3$-$\mathbf{g}_4$ space). The distribution of the four roots $x_i$ is shown (the numbers give the energy eigenvalue at that point), and we see that at the edges/corners 
some eigenvalues become degenerate.
\begin{figure}[hbtp]
\centering
\includegraphics[scale=0.4]{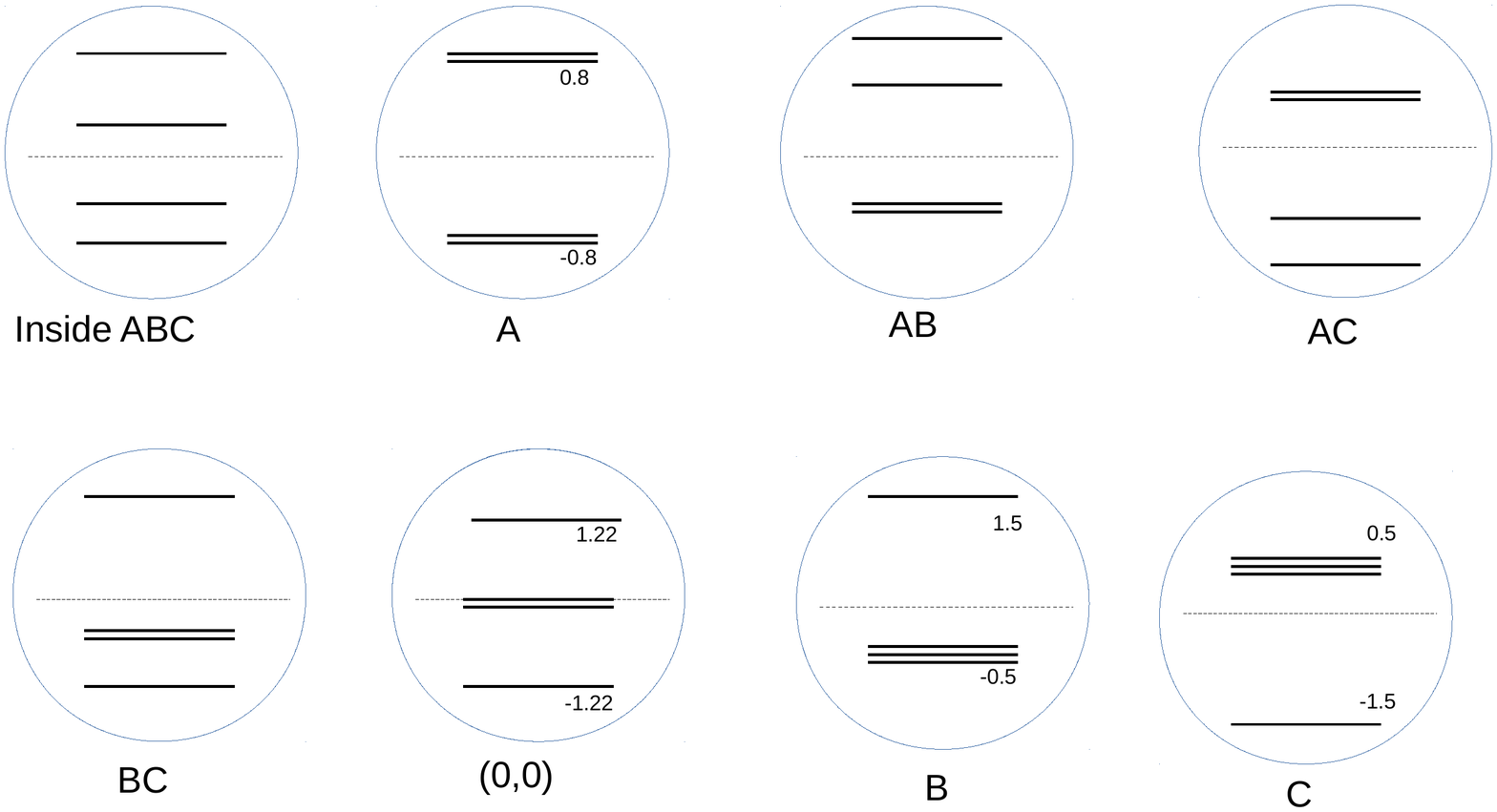}
\caption{\texttt{Degeneracy structure of Weyl 1-fermion states at different points in the gauge configuration space}}
\end{figure}
From (\ref{curveAB}), $\Phi_{bulk}$ diverges at the edge $AB$ where ground state degeneracy becomes 2, and also at the corner $B$, where degeneracy becomes 3.

At the edge $AB$, we find that (\ref{phiasP}) gives us
\begin{equation}
\Phi_{edge}=\frac{2}{9\mathbf{g}_2^2}\frac{x_1^2+3/4}{(x_1^2-1/4)^2}.
\label{1fphiedge}
\end{equation}
Approaching the corner $B$ from the bulk, the degeneracy changes from 1 to 3, whereas approaching it along $AB$, the degeneracy changes from 2 to 3. For the scalar potentials $\Phi_{bulk}$ and $\Phi_{corner}$, this implies
\[\Phi_{bulk}\rightarrow \frac{1}{3a^2}\frac{1}{(x_1+1/2)^2}; \quad\quad \Phi_{edge}\rightarrow \frac{2}{3a^2}\frac{1}{(x_1+1/2)^2}.\]
At this corner, we have $x_1=-1/2$, so both $\Phi_{bulk}$ and $\Phi_{edge}$ diverge.

However, we can also compute $\Phi_{corner}$ directly from (\ref{phiasP}). We find that
\begin{equation}
\Phi_{corner}=\frac{1}{2a^2}.
\end{equation}
We can also show that $\Phi_{bulk}$ diverges at the corner $A$ of Fig. 1 because the ground state degeneracy changes from 1 to 2, but $\Phi_{edge}$ does not, because the ground state degeneracy is unchanged.

Similarly, $\Phi_{bulk}$ is non-singular at the edges $AC$ and $BC$ (except at the point $C$ where it diverges, and the fermion degeneracy changes from 1 to 3).

For the 2-fermion ground state, the scalar potential can be obtained by using the 2-fermion Hamiltonian $H_f^{(2)}$ and taking the eigenstates (\ref{2fermionstate}). This is when the results of the 1-fermion calculation become particularly useful. It is clear that matrix elements of $H_f^{(2)}$ between two 2-fermion states are 0, unless these two states have atleast one 1-fermion state in common; the matrix element is then the matrix element of $H_f$ between the non-overlapping 1-fermion states.

\begin{figure}[hbtp]
\centering
\includegraphics[scale=0.4]{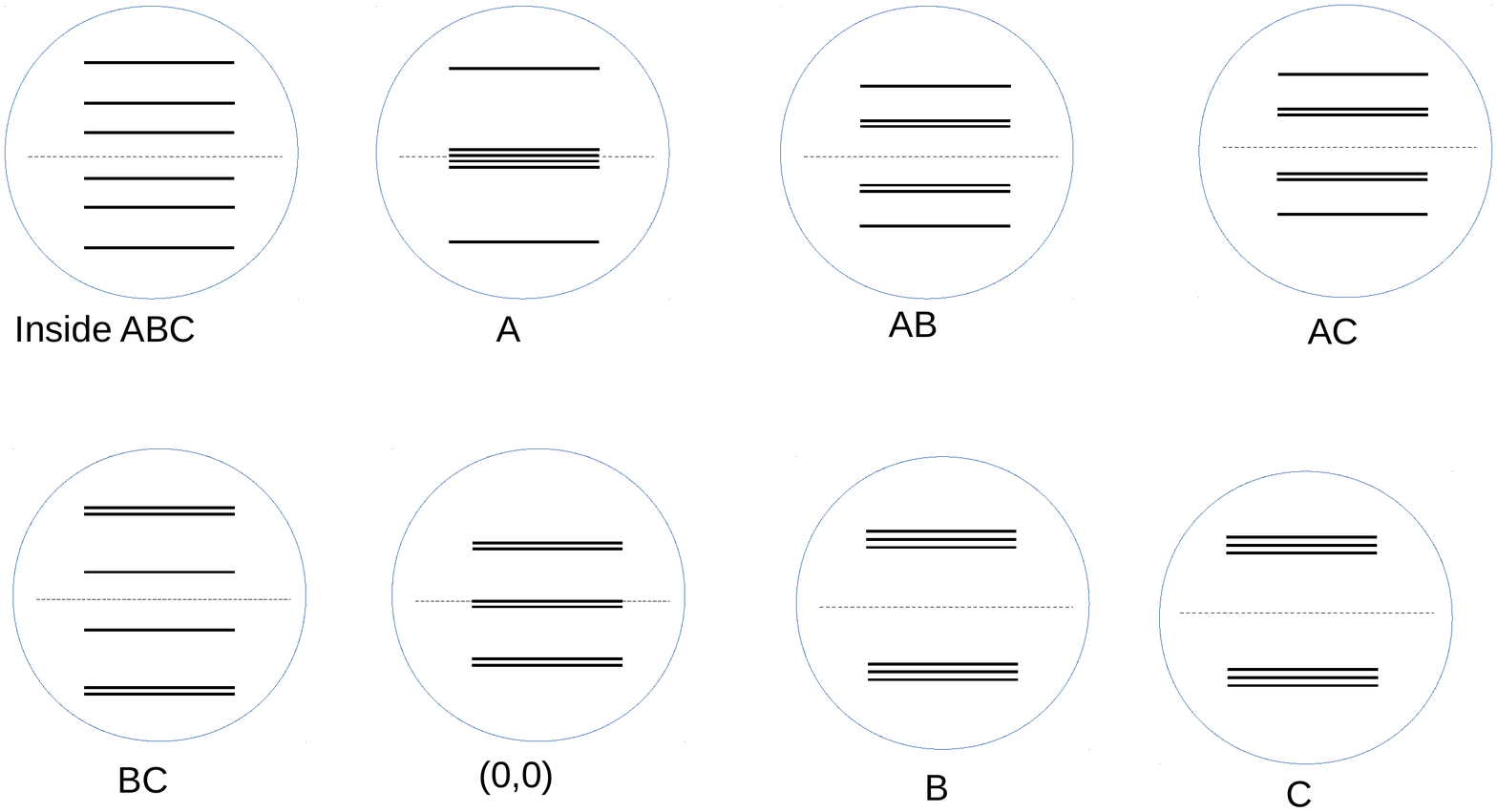}
\caption{\texttt{Degeneracy structure of Weyl 2-fermion states at different points in the gauge configuration space}}
\end{figure}
The scalar potential for the 2-fermion ground state can now be easily calculated, and turns out to be
\begin{equation}
\Phi^{(2)}_{bulk}=\frac{1}{(a_1-a_2)^2}+\frac{1}{(a_1+a_2)^2}+\frac{1}{(a_1-a_3)^2}+\frac{1}{(a_1+a_3)^2}
\end{equation}
or, in terms of dimensionless quantities $\mathbf{g}_i$,
\begin{equation}
\Phi^{(2)}_{bulk}=\frac{6}{\mathbf{g}_2^2}\frac{-y_1^6+5y_1^4+4(9/16-\mathbf{g}_4)(1-7y_1^2/3)}{(3y_1^4-6y_1^2+4(9/16-\mathbf{g}_4))^2}
\end{equation}
where $y_1 = y_1(\mathbf{g}_3,\mathbf{g}_4)$ is the smallest root of (\ref{2fchar}), the characteristic polynomial of the 2-fermion Hamiltonian.

Looking at Fig. 5, the ground state degeneracy changes from 1 in the bulk to 2 at the edge $BC$, and to 3 at the corners $B$ and $C$. Using (\ref{curveBC}), we see that $\Phi^{(2)}_{bulk}$ diverges as it approaches $BC$. At the corners $B$ and $C$, $y_1 = -1$, so $\Phi_{bulk}^{(2)}$ 
diverges here as well.

At the edge $AB$,
\begin{equation}
\Phi^{(2)}_{edge}=\frac{2}{9\mathbf{g}_2^2}\frac{9-6y_1^2+5y_1^4}{y_1^2(1-y_1^2)^2}.
\label{phi2edge}
\end{equation}
As one approaches the corner $B$,
\[\Phi^{(2)}_{edge}\rightarrow \frac{2}{9a^2}\frac{1}{(1+y_1)^2}.\]
Again because $y_1 = -1$ at $B$, $\Phi^{(2)}_{edge}$ diverges here.

At the corner $B$,
\begin{equation}
\Phi_{corner}^{(2)}=\frac{1}{a^2}.
\end{equation}

The scalar potential adds to the potential energy term in the effective Hamiltonian for the gauge fields:
\begin{eqnarray}
H_{eff}&=-\frac{g^2}{2}\mathcal{D}_{ia}\mathcal{D}_{ia}+V_{eff}, \\
V_{eff}&=\frac{1}{g^2}(V(M)+\epsilon_0)+\frac{g^2}{2}\Phi(M)
\end{eqnarray}

\begin{figure}[hbtp]
\centering
\includegraphics[scale=0.4]{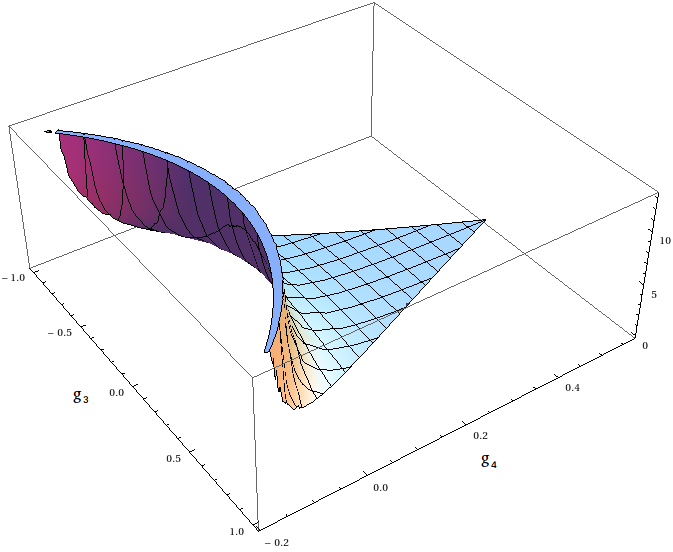}
\caption{$\Phi$ versus $\mathbf{g}_3$, $\mathbf{g}_4$}
\end{figure}

In $V_{eff}$, the $V(M)+\epsilon_0$ term is well-behaved everywhere. However, the scalar potential term $\Phi^{(2)}_{bulk}$ becomes singular along the edge $BC$, as seen in Fig. 6. For small $g$, $V(M)+\epsilon_0$ term dominates over $\Phi^{(2)}_{bulk}$ almost everywhere in the gauge configuration space, except at the places where the latter becomes singular.


\section{Quantum Dynamics of Gauge Fields}

We are now equipped to study the dynamics of the gauge fields, as dictated by the effective Hamiltonian (\ref{Heff}). It would be an interesting 
problem in itself to study the complete effective Hamiltonian and in particular questions about possible self-adjoint extensions, corresponding spectrum and so on. Rather than embark on this difficult functional analytic problem, we will argue that a study of the singularity structure of the effective scalar potential provides us with information on the quantum phase structure of this theory. 

Different strata of ${\cal C}$ are characterized by properties of the singular values $a_i$. Changing the $a_i$'s can take us from one stratum to another, while gauge and physical rotations leave us on the same stratum. Similarly, the fermion spectrum depends only on $a_i$ or equivalently $\mathbf{g}_i$. It thus suffices to ignore physical 
rotations,  quotient by gauge transformations, and study the dynamics of the gauge field only in the 
$\mathbf{g}_2$-$\mathbf{g}_3$-$\mathbf{g}_4$ 
space or equivalently, the space of $a_i$'s.

This gives us an important simplification: the adiabatic connection does not contribute to the effective Hamiltonian, since there is no component of 
$\mathcal{A}$ purely in the directions of the singular values. Then
\begin{equation}
H_{eff}=\frac{g^2}{2}\left(-\Delta+\Phi\right)+\frac{1}{g^2}V(M)+\epsilon_0(M)
\end{equation}
where $\Delta$ denotes the Laplacian operator. It can be explicitly evaluated by noting that we have a natural metric on the space of $M_{ia}$:
\begin{equation}
ds^2={\rm Tr}(dM^TdM)
\end{equation}

Iwai \cite{iwai} has studied the metric and the Laplacian in SVD variables, and we will draw extensively from his results. In SVD, the metric becomes 
\begin{equation}
ds^2=\sum_ida_i^2+\sum_{i}\sum_{j\neq i} a_i^2(\omega_{jR}^2+\omega_{jS}^2)-2 \sum_{i}\sum_{j\neq i}\sum_{k\neq i,k\neq j}\omega_{iR}\omega_{iS} a_j a_k
\label{metric}
\end{equation}
where $R^T dR=-i \omega_{iR}T_i$ and $S^TdS= -i \omega_a^S T_S$ denote the left-invariant one-forms on the $SO(3)_R$ and $SO(3)_S$ respectively. Thus the metric is
\begin{equation}
g_{IJ}=\left(\begin{array}{ccc}
g_{aa} & 0 & 0 \\ 
0 & g_{RR} & g_{RS} \\ 
0 & g_{SR} & g_{SS} \\
\end{array} \right)
\end{equation}
where
\[
g_{aa}=\mathbb{I}_3; \quad
g_{RR}=g_{SS}=\left(\begin{array}{ccc}
a_2^2+a_3^2 & 0 & 0 \\ 
0& a_1^2+a_3^2 & 0  \\ 
0& 0& a_1^2+a_2^2  \\
\end{array} \right) ; \quad g_{RS}=g_{SR}=\left(\begin{array}{ccc}
-2a_2a_3 & 0 & 0 \\ 
0& -2a_1a_3 & 0  \\ 
0& 0& -2a_1a_2  \\
\end{array} \right).\]
The Laplacian can be calculated by using
\begin{equation}
\Delta f=\frac{1}{\sqrt{|\det g|}}\partial_I\left(\sqrt{|\det g|}g^{IJ}\partial_J f\right).
\label{Laplacian}
\end{equation}
In the bulk, when the singular values are all different, the Laplacian takes the form
\begin{align}
\Delta &= \nonumber \frac{\partial^2}{\partial a_1^2}+\frac{\partial^2}{\partial a_2^2}+\frac{\partial^2}{\partial a_3^2}\\
&\nonumber+2a_1\left(\frac{1}{a_1^2-a_2^2}+\frac{1}{a_1^2-a_3^2}\right)\frac{\partial}{\partial a_1}+2a_2\left(\frac{1}{a_2^2-a_3^2}+\frac{1}{a_2^2-a_1^2}\right)\frac{\partial}{\partial a_2}+2a_3\left(\frac{1}{a_3^2-a_1^2}+\frac{1}{a_3^2-a_2^2}\right)\frac{\partial}{\partial a_3}\\
&\nonumber - \frac{a_2^2+a_3^2}{(a_2^2-a_3^2)^2}(L_{1r}^2+L_{1S}^2)-\frac{a_3^2+a_1^2}{(a_3^2-a_1^2)^2}(L_{2r}^2+L_{2S}^2)-\frac{a_1^2+a_2^2}{(a_1^2-a_2^2)^2}(L_{3r}^2+L_{3S}^2)\\
&-\frac{4a_2a_3}{(a_2^2-a_3^2)^2}L_{1R}L_{1S}-\frac{4a_1a_3}{(a_1^2-a_3^2)^2}L_{2R}L_{2S}-\frac{4a_1a_2}{(a_1^2-a_2^2)^2}L_{3R}L_{3S}
\end{align}
where $L^i_{R}$, and $L^i_{S}$ are the left invariant vector fields on $SO(3)_R$ and $SO(3)_S$ respectively, and
\begin{equation}
\left(\begin{array}{c}
	L_{iR} \\
	L_{iS}
	\end{array}\right) = \left(\begin{array}{cc}
						g_{RR} & g_{RS} \\
						g_{SR} & g_{SS}
					     \end{array}\right) \left(\begin{array}{c}
					     					\omega_{iR}\\
										\omega_{iS}
									      \end{array}\right).
\end{equation}

Restricting only to variations in $a_i$, the Laplacian in the bulk is given by
\begin{align}
\Delta_{bulk} &= \nonumber \frac{\partial^2}{\partial a_1^2}+\frac{\partial^2}{\partial a_2^2}+\frac{\partial^2}{\partial a_3^2}\\
&+2a_1\left(\frac{1}{a_1^2-a_2^2}+\frac{1}{a_1^2-a_3^2}\right)\frac{\partial}{\partial a_1}+2a_2\left(\frac{1}{a_2^2-a_3^2}+\frac{1}{a_2^2-a_1^2}\right)\frac{\partial}{\partial a_2}+2a_3\left(\frac{1}{a_3^2-a_1^2}+\frac{1}{a_3^2-a_2^2}\right)\frac{\partial}{\partial a_3}
\end{align}
Then
\begin{equation}
H_{bulk}=\frac{g^2}{2}\left(-\Delta_{bulk}+\Phi_{bulk}\right)+\frac{1}{g^2}V(M)+\epsilon_0(M).
\end{equation}
The apparent divergence of the Laplacian at coincident singular values may seem cause for concern, but as Iwai argues, the volume factor $\sqrt{\det g} = (a_1^2-a_2^2)(a_1^2-a_3^2)(a_2^2-a_3^2)$ in SVD coordinates makes 
the contribution of the kinetic term to the energy integral $\langle \Psi, -\Delta \Psi\rangle$ finite. This is also borne out by the fact that in standard cartesian coordinates $M_{ia}$, the Laplacian of the metric Tr 
$dM^T dM$ has no singularities and is essentially self-adjoint.

Recall that $V(M)$ and $\epsilon(M)$ are well-behaved everywhere but $\Phi_{bulk}$ becomes singular along the edge $BC$, and corners $B$ and $C$. Finiteness of energy requires that the domain of $H_{bulk}$ contain only functions that vanish as one approaches $BC$, and $B$ and $C$.

What is the dynamics of the gauge fields restricted to the edge $BC$, where $a_1=a_2=a$? On this edge, the metric (\ref{metric}) takes the form:
\begin{equation}
ds^2= 2da^2+da_3^2+(\omega_{1R}^2+\omega_{1S}^2+\omega_{2R}^2+\omega_{2S}^2)(a^2+a_3^2)-4a a_3(\omega_{1R}\omega_{1S}+\omega_{2R}\omega_{2S})+2a^2(\omega_{3R}-\omega_{3S})^2.
\end{equation}

Now the six $\omega$'s are not all independent, as is evident from the fact that the volume factor $(a_1^2-a_2^2)(a_1^2-a_3^2)(a_2^2-a_3^2)$ vanishes at this edge. We can understand this better by noting that a rotation about the third axis commutes with the matrix of singular values $A=\text{diag}(a,a,a_3)$. So if we separate $R$ as $R=R_1 R_2 R_3$, a product of different rotations about the three axes, and similarly $S=S_1 S_2 S_3$, then the matrices $R_3$ and $S_3$ combine to give a single rotation matrix:
\[ M= (R_1 R_2 R_3)A(S_3^TS_2^TS_1^T)= (R_1 R_2)A( (R_3 S_3^T) S_2^T S_1^T).
\]
So there are only 5 angular coordinates that parametrize $M$, as opposed to $6$, since the left and right rotations about the third axis can be combined. As expected, the angular momenta about the third axis are equal and opposite:
\begin{equation}
L_{3R}=-L_{3S}=2a^2(\omega_{3R}-\omega_{3S})^2.
\end{equation}
In this case, we can take our independent coordinates as $a,a_3, \omega_{1R},\omega_{2R},\omega_{1S},\omega_{2S}$ and $\omega_{3R}-\omega_{3S}$, and obtain the metric
\begin{equation}
g_{IJ}=\left(\begin{array}{cc}
g_{aa} & 0\\
0 & g_{\omega \omega} 
\end{array} \right)
\end{equation}

where
\[
g_{aa}=\left(\begin{array}{cc}
2 & 0\\
0 & 1
\end{array} \right)
\]

and

\[
g_{\omega\omega}=\left(\begin{array}{ccccc}
a^2+a_3^2 & 0 & -2aa_3 & 0 & 0 \\ 
0& a^2+a_3^2 &  0& -2aa_3 & 0 \\ 
-2a a_3 & 0 & a^2+a_3^2 & 0 & 0 \\ 
0& -2aa_3 & 0 & a^2+a_3^2 & 0 \\
0& 0& 0& 0& 2a^2  \\
\end{array} \right).\]
So the volume factor becomes $\sqrt{\det g} = \sqrt{2}a(a^2-a_3^2)^2$. Using (\ref{Laplacian}) and restricting to the dynamics along the directions of the singular values, we find that the Laplacian along the edge $BC$ is
\begin{equation}
\Delta_{edge}=\frac{\partial^2}{\partial a^2}+\frac{\partial^2}{\partial a_3^2}+\frac{1}{2a}\frac{\partial}{\partial a}+\frac{1}{(a^2-a_3^2)}\left( a\frac{\partial}{\partial a}-a_3\frac{\partial}{\partial a_3}\right),
\end{equation}
and the corresponding $H_{eff}$ is
\begin{equation}
H_{edge}=\frac{g^2}{2}\left(-\Delta_{edge}+\Phi^{(2)}_{edge}\right)+\frac{1}{g^2}V(a,a_3)+\epsilon_0(a,a_3).
\end{equation}
Again, there is an apparent divergence of the Laplacian as $a_3\rightarrow a$, but the volume factor $\sqrt{2}a(a^2-a_3^2)^2$ ensures that contribution of the kinetic energy term to the energy integral is finite. The potential $\Phi^{(2)}_{edge}$ however becomes singular when $a=a_3$, i.e., as one approaches the corner $B$. So finiteness of energy requires that wavefunctions corresponding to $H_{edge}$ vanish as one approaches the corner.

Wavefunctions in the domain of $H_{edge}$ cannot be constructed from the wavefunctions in the domain of $H_{bulk}$ as the latter vanish at the edge. Thus we find that the full Hilbert space the describes the dynamics of the gauge field has superselection sectors, one sector describing the dynamics in the bulk, another describing dynamics at the edge. These sectors cannot mix.

Our discussion in Section 4.3 tells us that when two singular values coincide, the fermion Hamiltonian possesses an extra physical symmetry corresponding 
to rotations about one of the axes. We may therefore think of this phase as one in which gauge- and physical rotations along this particular axis 
are locked.

Now let us study the dynamics of the gauge field configuration restricted to the corner $B$, where $a_1=a_2=a_3=a$. The corner $C$ is related to the corner $B$ via a parity transformation, and is equivalent to $B$. At $B$, the matrix $M=a G, G \in SO(3)$, so the metric Tr $dM^TdM$ takes the form
\begin{equation}
ds^2=3 da^2 - a^2 {\rm Tr} (G^T dG)^2
\end{equation}
Restricting to the dynamics purely along the $a$-direction, we find that Laplacian is 
\begin{equation}
\Delta_{corner}=\frac{1}{3}\left(\frac{\partial^2}{\partial a^2}+\frac{3}{a}\frac{\partial}{\partial a}\right).
\end{equation}
Thus the effective Hamiltonian is
\begin{equation}
H_{corner}=\frac{g^2}{2}\left(-\Delta_{corner}+\Phi_{corner}\right)+\frac{1}{g^2}
V(a\mathbf{I})+\epsilon_0(\mathbf{I})
\end{equation}
where $\Phi_{corner}=\frac{1}{2a^2}$.

It is not difficult to show that $H_{corner}$ is essentially self-adjoint. For any $\psi$ in the domain of $H_{corner}$, we perform the transformation $\psi\rightarrow U \psi = a^{-3/2} \psi$. The Hamiltonian transforms as $H_{corner} \rightarrow U H_{corner} U^{\dagger}$, giving
\begin{eqnarray}
H_{corner} &=& \frac{g^2}{2}\frac{1}{3}\left(-\frac{\partial^2}{\partial a^2} + \frac{3}{4a^2}\right) + \frac{g^2}{2} \Phi_{corner} + \frac{1}{g^2} V(a\mathbf{I})+\epsilon_0(\mathbf{I}) , \\
&=& \frac{g^2}{2}\frac{1}{3}\left(-\frac{\partial^2}{\partial a^2} + \frac{9}{4a^2}\right) + \frac{1}{g^2} V(a\mathbf{I})+\epsilon_0(\mathbf{I}).
\end{eqnarray}
The operator 
\begin{equation}
-\frac{\partial^2}{\partial a^2} + \frac{\alpha}{a^2}
\end{equation}
has been extensively studied \cite{meetz},\cite{narnhofer}, and is known to be essentially self-adjoint for $\alpha>3/4$, which is our situation. This means that no boundary conditions are necessary at $a=0$.

Wavefunctions belonging to the domain of $H_{corner}$ cannot be constructed out of those belonging to both the domains of $H_{edge}$ and $H_{bulk}$, since they vanish at the corners. This characterizes another superselection sector. As we saw in Section 4.3, this situation corresponds to having an enhanced $SO(3)$ symmetry for the fermion Hamiltonian. The color- and physical rotations are locked into a single $SO(3)$. This particular phase is thus the analog of the color-spin locked phase that has been discussed in the context of 3-color QCD \cite{Schafer:2000tw}. 

Thus there are different effective Hamiltonians governing the dynamics of the Yang-Mills in the three different sectors, namely 
the bulk, the edge $BC$, and the corner $B$ (and equivalently $C$). These Hamiltonians have different domains, and there is 
no observable connecting one sector to another. These superselection sectors can be 
interpreted as three different quantum phases.

\section{The Matrix Model coupled to Dirac Fermions}

A Dirac fermion is made up of a left Weyl fermion and a right Weyl fermion. 
The Lagrangian for the Yang-Mills Matrix Model coupled to Dirac fermions is simply
\begin{equation}
L=-\frac{1}{4g^2}F_{\mu\nu}^aF^{a\mu\nu}
+\frac{1}{g^2}\left(i\bar{\psi}_{A} \gamma^\mu (\mathcal{D}_\mu\psi)_{A}
-m\bar{\psi}\psi+\bar{\psi}\gamma^0\psi\right)
\label{LYMD}
\end{equation}
where $\bar{\psi}=\psi^\dagger\gamma^0$ and the gamma matrices in the Weyl representation are
\begin{equation}
\gamma^\mu=\left(\begin{array}{cccc}
0 & \sigma^\mu \\ 
\bar{\sigma}^\mu & 0  
\end{array}  \right).
\end{equation}
Again, we have rescaled $\psi\rightarrow g \psi$ to write the Lagrangian in the form (\ref{LYMD}).

Quantizing the theory in the Born-Oppenheimer approximation as before, we would obtain the Hamiltonian
\begin{equation}
H=H_{YM}+H_f
\end{equation}
where
\begin{equation}
H_f=\psi^\dagger_{\alpha A}(H^f)_{\alpha A,\beta B}\psi_{\beta B}
\end{equation}
\begin{equation}
(H^f)_{\alpha A,\beta B}=-\delta_{AB}\delta_{\alpha\beta}+\frac{1}{2}(\tau_c)_{AB}(\gamma^0\gamma^i)_{\alpha\beta}M_{ic}+m\delta_{AB}(\gamma^0)_{\alpha\beta}.
\end{equation}

Taking the 1-fermion states to be of the form
\begin{equation}
|\psi^{(1)}_n\rangle = D^n_{\alpha A}\psi^\dagger_{\alpha A}|0\rangle
\end{equation}
we find that the $D_{\alpha A}$'s obey
\begin{equation}
(H_f)_{\alpha A, \beta B} D^n_{\beta B} = \epsilon_n D^n_{\alpha A}.
\end{equation}
Since $\psi$ is a 4-component Dirac fermion, $\alpha$ takes 4 values. In the 2-component notation,
\begin{equation}
\psi=\left(\begin{array}{cc}
\psi_L\\
\psi_R
\end{array}  \right)
\end{equation}
and
\begin{equation}
H^f=\left(\begin{array}{cccc}
H_1 & m \\ 
m & -H_1  
\end{array}  \right)
\end{equation}
where $H_1$ denotes the Hamiltonian (\ref{Hf}) for the single chirality fermion.

In this article, we shall discuss only the $m=0$ case.

\subsection{Spectrum of the Dirac Hamiltonian}

If mass $m=0$, then

\begin{equation}
H^f=\left(\begin{array}{cccc}
H_1 & 0 \\ 
0 & -H_1  
\end{array}  \right).
\end{equation}
Then the spectrum is of the form:
\begin{equation}
D^n_+=\left(\begin{array}{cc}
c^n\\
0
\end{array}  \right);\quad eigenvalue\hspace{1 mm}+\epsilon_n
\end{equation}
\begin{equation}
D^n_-=\left(\begin{array}{cc}
0\\
c^n
\end{array}  \right);\quad eigenvalue\hspace{1 mm}-\epsilon_n
\end{equation}
where $c^n$ are the single chirality eigenstates and $\epsilon_n$ are the eigenvalues for the single chirality.

In terms of the left and right Weyl components, the Hamiltonian can be written as:

\begin{equation}
H_f=\psi_L^\dagger H_1\psi_L-\psi_R^\dagger H_1\psi_R
\end{equation}
where $\psi_L^\dagger$ creates a left-handed fermion and $\psi_R^\dagger$ creates a right-handed fermion.

The eigenstates are:-
\begin{equation}
|n;L\rangle=c^n_{\alpha A}(\psi_L)_{\alpha A}^\dagger|0\rangle;\quad H_f|n;L\rangle=\epsilon_n|n;L\rangle
\end{equation}
\begin{equation}
|n;R\rangle=c^n_{\alpha A}(\psi_R)_{\alpha A}^\dagger|0\rangle;\quad H_f|n;R\rangle=-\epsilon_n|n;R\rangle
\end{equation}

Multiparticle states can be formed by taking tensor products of one-particle states. A state with $m$ left-handed fermions and $n$ right-handed fermions is represented by
\begin{equation}
|l_1,...l_m;r_1...r_n\rangle=\prod_{i=1}^m\prod_{j=1}^n (c^{l_i}\psi_L^\dagger) (c^{r_j}\psi_R^\dagger)|0\rangle
\end{equation}
with energy eigenvalue
\begin{equation}
\epsilon=\sum_{l=1}^m \epsilon_l-\sum_{r=1}^n \epsilon_r.
\end{equation}
In particular we want to examine the 2-fermion states. We can put two $L-$type, two $R-$type, or 1 of each type of fermions. The spectrum can be easily obtained via $(5.16)$, and is tabulated below.\\


\begin{tabular}{|c||c|c|c|}
\hline 
Type     & LL & RR &LR \\ 
\hline \hline
		&			&			&$\pm (a_1+a_2)$\\  	
   		& $\pm a_1$ 	& $\pm a_1$     & $\pm (a_1-a_2)$\\ 
Energy 	& $\pm a_2$ 	& $\pm a_2$	& $\pm (a_1+a_3)$\\  
   	    	& $\pm a_3$ 	& $\pm a_3$	& $\pm (a_1-a_3)$\\  
		&			&			& $\pm (a_2+a_3)$\\ 
		&			&			& $\pm (a_2-a_3)$\\ 
		&			&			& $0,0,0,0$\\
\hline 
\end{tabular}
\\

The $LL$ and the $RR$ energy levels are exactly the energy levels of the 2-fermion states for the Weyl fermions. Each of them is doubly degenerate. But there is an extra set of energy levels coming from putting two fermions of opposite chirality.

Note that the ground state, with energy $-(a_1+a_2)$ is of the type $LR$, so the ground state of the theory does not have well-defined chiral symmetry.

\subsection{Scalar Potential for massless Dirac Fermions}

Since all the computations for the Dirac case are exactly analogous to our earlier discussion, we will be brief here.

The Dirac Hamiltonian is diagonal in the $L$-$R$ basis, so matrix elements of $\partial_{ia}H$ are zero between an $L$- and an $R$- state. In the calculation of the scalar potential for a particular state, the only contribution comes from other states of the same handedness. This is equivalent to working with a single Weyl fermion with fixed handedness. The expression for the scalar potential of one-particle massless Dirac fermions is the 
same as that for the single Weyl fermion:
\begin{equation}
\Phi_{D; 1-fermion}=\Phi_{W;1-fermion}.
\end{equation}

The case of interest is the $2$-fermion sector, which is anomaly-free. Again the scalar potential can be calculated by taking $H_f^{(2)}=H_f\otimes \mathbf{1}+\mathbf{1}\otimes H_f$. Now the $2$-fermion ground state is $LR$-type; so only other $LR$-states will contribute to the scalar potential. Since, the ground state wave-function is
\begin{equation}
|gs\rangle=c^1_{\alpha A}c^4_{\beta B}(\psi^L)^\dagger_{\alpha A}(\psi^R)^\dagger_{\beta B}|0\rangle,
\end{equation}
the scalar potential for the $2$-fermion ground state works out to be 
\begin{equation}
\Phi_{D;2-fermion}(gs)=\Phi_{W;1-fermion}(x_1)+\Phi_{W;1-fermion}(x_4)
\label{phidirac}
\end{equation}
where $x_4$ and $x_1$ are the highest and lowest roots of the polynomial $(\ref{cheq})$. This formula correctly takes into account the change in degeneracy of the $1$-fermion Weyl states. Indeed by looking at Fig. 7, one can see that degeneracy changes from $1$ to $2$ in going from the bulk to the edge $AB$ (or $AC$, ), to $3$ at the corner $B$ (or $C$), and becomes $4$ at the corner $A$. Under parity transformation, the  edge $AC$ is related to $AB$, and the corner $C$ to $B$.

\begin{figure}[hbtp]
\centering
\includegraphics[scale=0.4]{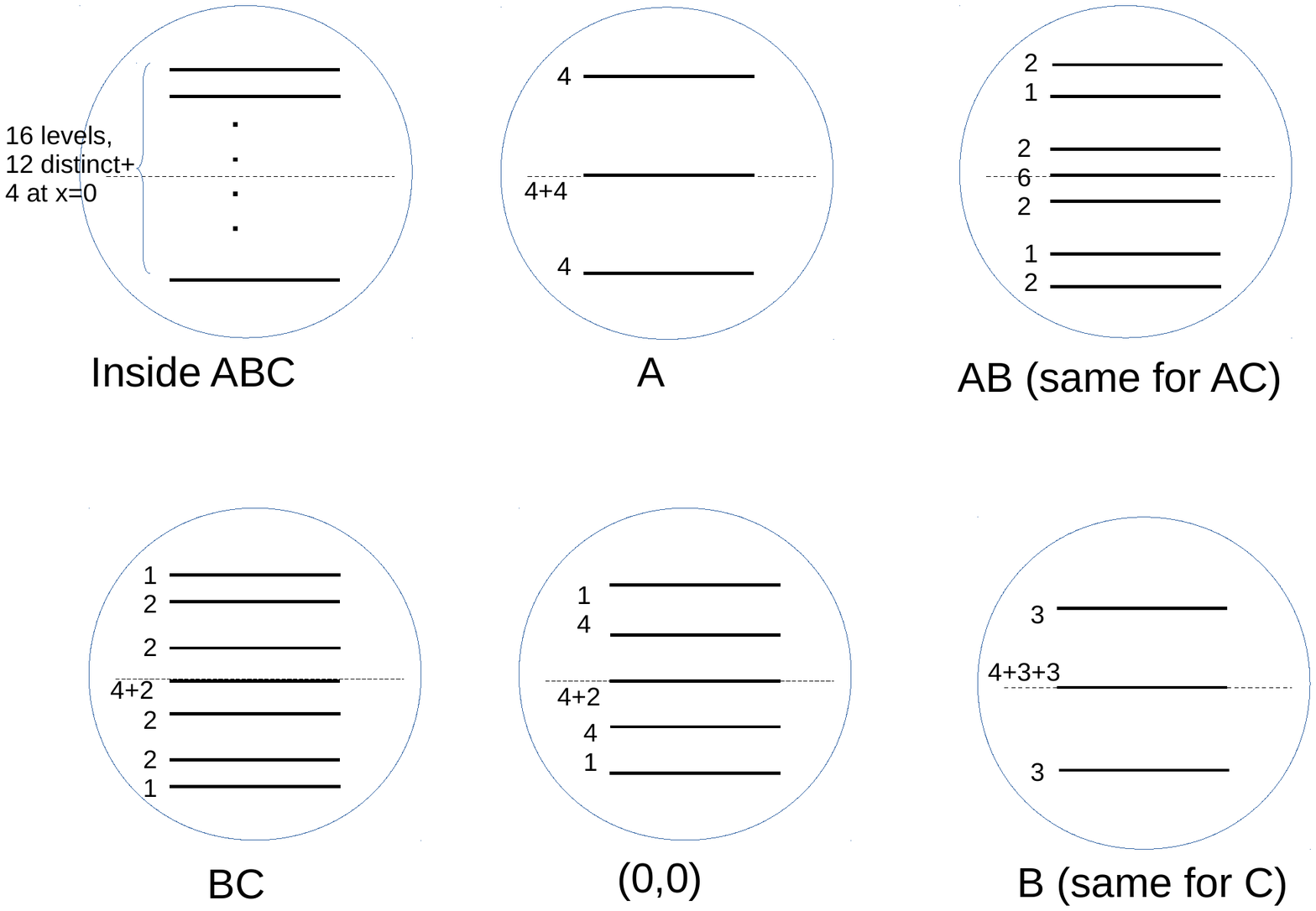}
\caption{\texttt{Degeneracy structure for LR states (the numbers denote degeneracy)}}
\end{figure}

Eq. (\ref{phidirac}) can be applied to find the scalar potential in the bulk, as well as at these edges and corners, making sure one uses the correct expression for the scalar potential for each Weyl component. For example, on the edge $AB$, the state corresponding to $x_1$ is doubly degenerate and so the corresponding projector is rank $2$, while the state corresponding to $x_4$ is non-degenerate and the projector is rank one. So in the expression for the scalar potential on $AB$, we must use (\ref{1fphiedge}) for $x_1$, and for $x_4$ we must use (\ref{1fphibulk}) evaluated at 
the edge $AB$.
 
By an analysis similar to section 7, it can be argued that the effective theory for $LR$ ground state of the Dirac fermion has four different superselection sectors: the bulk, the edge $AB$ (or $AC$), the corner $B$ (or $C$), and the corner $A$. The scalar potential evaluated for each sector shows singular behaviour as we approach the other sectors, requiring that the wavefunctions vanish there, and resulting in distinct domains for the effective Hamiltonian in the different sectors. Thus this theory has four different quantum phases.

Fig. 8 shows that $\Phi$ is singular along the edges $AB$ and $AC$.
\begin{figure}[hbtp]
\caption{Plot of $\Phi$ for the $LR$ levels}
\centering
\includegraphics[scale=0.5]{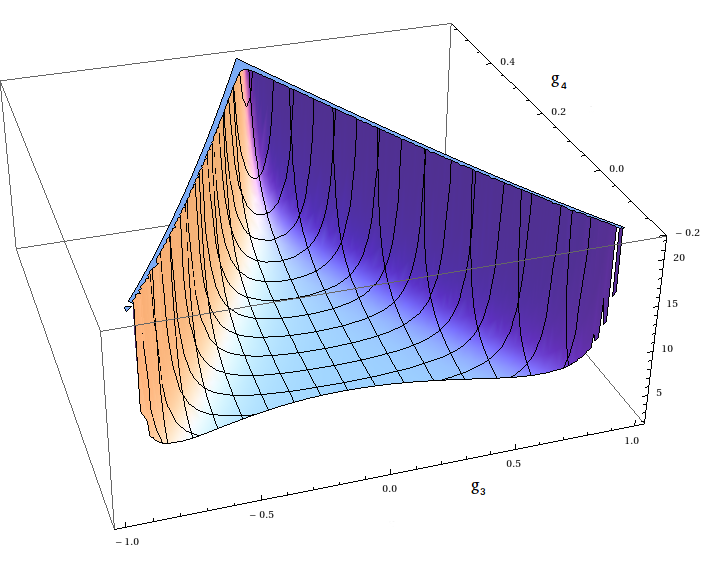}
\end{figure}


The $LL$ and $RR$ states are equivalent to the $2$-fermion Weyl fermion states, so the scalar potential calculations for them are the same as in Section 6. For $SU(2)$ gauge theory these are baryons, while the $LR$ states are mesons.

We can add more fermions, making sure that the fermion number is even, and the scalar potentials will add up appropriately. One can look at this as an introduction of a chemical potential $\mu$ to shift the Fermi surface, so we can fill all states below the zero of the chemical potential. This provides an interesting interpretation of our phases. For example for the case of a single Weyl fermion, if we look at the theory in the bulk and choose the chemical potential such that it is just above the lowest energy level, then the theory is anomalous; however the places where the ground state becomes degenerate are free from the anomaly, since there are two levels that need be filled. This happens when $\mu$ is greater than some critical value $\mu_c$. So for a given value of $\mu$, only certain phases exist in the theory. 

Similar considerations apply for the Dirac theory as well. Now the $L$ and $R$ levels are degenerate. If we choose $\mu$ such that it is just above the lowest energy level, the only possibility in the bulk is to have an $LR$ state, i.e. a meson. But when the ground state becomes degenerate, we have more choices to fill the energy levels. In this phase, $LL$, $RR$ and $LR$ type states can all exist, i.e. we have both baryons and mesons. This is reminiscent of the discussion in \cite{Kogut:2000ek}, where diquark condensates appear when $\mu$ is larger than some critical value.
 
%

\section{Conclusions}
Yang-Mills theory coupled to fermions has a rich phase structure, and by studying the matrix model limit of 
the theory, we have been able to deduce several of its aspects. Much of this information comes from quantizing 
the theory in the adiabatic approximation, by treating the fermionic variables as fast degrees, and the gauge field
as the slow variable. This Born-Oppenheimer quantization induces both an adiabatic connection as well as an 
effective scalar potential for the gauge Hamiltonian. The adiabatic connection has simple but interesting implications 
for $SU(2)$ gauge invariance (or its breaking), as we see from the existence of the color-spin locked phase. The 
effective scalar potential has a sophisticated singularity structure, coinciding with locations where the fermion 
degeneracy changes. The effective potential is responsible for creating superselection sectors in the gauge 
field Hilbert space. We suggest that these different superselection sectors be interpreted as different 
quantum phases of the theory.

For the case of $SU(2)$ gauge field coupled to two left- (or right-) handed fermions, we find that there are {\it four} 
different quantum phases, while for the case of a massless Dirac fermion, we find that there are {\it six}. Of these 
we are able to identify one phase, the color-spin locked phase, as one known in literature. The other phases, to 
our knowledge, seem to be new.

Substantial progress for the case of $SU(2)$ has been possible because of the availability of SVD 
for $3\times3$ matrices $M_{ia}$. The physically more interesting and relevant case is the $SU(3)$ matrix model 
coupled to (almost) massless quarks. The matrix model is now based on a $3 \times 8$ rectangular matrix $M_{ia}$, 
and SVD is not appropriate since it does not correctly capture the gauge symmetry of the model. For unambiguous identification of phases, we expect that new theoretical techniques will be needed.

\noindent {\bf Acknowledgments:} 
It is a pleasure to thank N. Acharyya, M. Asorey, A. P. Balachandran, V. Errasti Diez, S. Mukherjee, D. Sen and 
V. Shenoy for discussions.


\begin{thebibliography}{99}

\bibitem{Alford:2007xm} 
  M.~G.~Alford, A.~Schmitt, K.~Rajagopal and T.~Sch\"afer,
  Rev.\ Mod.\ Phys.\  {\bf 80}, 1455 (2008)
  [arXiv:0709.4635 [hep-ph]].


\bibitem{Balachandran:2014iya}
  A.~P.~Balachandran, S.~Vaidya and A.~R.~de Queiroz,
  Mod.\ Phys.\ Lett.\ A {\bf 30} (2015) no.16,  1550080

\bibitem{Balachandran:2014voa}
  A.~P.~Balachandran, A.~de Queiroz and S.~Vaidya,
  Int.\ J.\ Mod.\ Phys.\ A {\bf 30} (2015) no.09,  1550064
 
  
\bibitem{Singer:1978dk} 
  I.~M.~Singer,
  Commun.\ Math.\ Phys.\  {\bf 60}, 7 (1978).
  
\bibitem{Narasimhan:1979kf}
  M.~S.~Narasimhan and T.~R.~Ramadas,
  Commun.\ Math.\ Phys.\  {\bf 67} (1979) 121.

\bibitem{berry}
M.V. Berry, in \textit{Geometric Phases in Physics}, edited by A. Shapere and F. Wilczek (World Scientific, Singapore,1990), p. 14.

\bibitem{Bohm}
 A. Bohm, B. Kendrick, M~E.~Loewe, L.~J.~Boya,
 J. Math. Phys. 33, 977 (1992); 

  
\bibitem{Schafer:2000tw} 
  T.~Sch\"afer,
  Phys.\ Rev.\ D {\bf 62}, 094007 (2000) 5  
  

\bibitem{Sachdev}
S. Sachdev,
\textit{Quantum Phase Transitions}, 
Cambridge University Press, Cambridge, 2011.

\bibitem{Zanardi}
A. T. Rezakhani, D. F. Abasto, D. A. Lidar, P. Zanardi,  
Phys. Rev. A 82, 012321 (2010);

\bibitem{Dutta}
A. Dutta, G. Aeppli, B. K. Chakrabarti, U. Divakaran, T. F. Rosenbaum, D. Sen,
\textit{Quantum Phase transitions in Transverse Field Spin Models: From Statistical Physics to Quantum Information}, 
Cambridge University Press, Cambridge, 2015.

\bibitem{Sen:1985dc} 
  D.~Sen,
  J.\ Math.\ Phys.\  {\bf 27}, 472 (1986).

\bibitem{kato}
T. Kato,
\textit{PerturbationTheory for Linear Operators},
Springer-Verlag, Berlin, 1995.

\bibitem{pv1}
M. Pandey and S. Vaidya,
{\it in preparation}.

\bibitem{Witten:1982fp}
  E.~Witten,
  Phys.\ Lett.\ B {\bf 117} (1982) 324.
  
\bibitem{quartic}
E. L. Rees,
Amer. Math. Monthly 29, 2 (1922). 

\bibitem{prasolov}
V. V. Prasolov,
\textit{Polynomials},
Springer-Verlag, Berlin, 2010.

\bibitem{iwai}
T. Iwai,
J. Phys. A: Math. Theor. {\bf 43} 415204 (2010).

\bibitem{meetz}
K. Meetz,
Nuovo Cimento {\bf 34}: 690 (1964).

\bibitem{narnhofer}
H. Narnhofer,
Acta Physica Austriaca {\bf 40}: 306Ð322 (1974).

\bibitem{Kogut:2000ek} 
  J.~B.~Kogut, M.~A.~Stephanov, D.~Toublan, J.~J.~M.~Verbaarschot and A.~Zhitnitsky,
  Nucl.\ Phys.\ B {\bf 582}, 477 (2000)



\end{thebibliography}
\end{document}